\documentclass[aps,showpacs,preprintnumbers,amsmath,amssymb,twocolumn]{revtex4}
\UseRawInputEncoding
\usepackage{dcolumn}
\usepackage[dvips]{epsfig}
\usepackage{float}
\usepackage{graphics,epsfig}
\usepackage{graphicx}
\usepackage{dcolumn}
\usepackage{bm}
\usepackage{gensymb}
\usepackage{footnote}
\usepackage{booktabs}
\usepackage{subfigure}

\begin{document}
\baselineskip=0.5 cm
\title{\bf Chaotic Shadows of Black Holes: A Short Review}

\author{ Mingzhi Wang$^{1}$\footnote{wmz9085@126.com}, Songbai Chen$^{2,3}$\footnote{Corresponding author: csb3752@hunnu.edu.cn}, Jiliang Jing$^{2,3}$\footnote{jljing@hunnu.edu.cn}
}
\affiliation{$ ^1$School of Mathematics and Physics, Qingdao University of Science and Technology, Qingdao, Shandong 266061, People's Republic of China \\
$ ^2$Institute of Physics and Department of Physics, Key Laboratory of Low Dimensional Quantum Structures
and Quantum Control of Ministry of Education, Synergetic Innovation Center for Quantum Effects and Applications,
Hunan Normal University, Changsha, Hunan 410081, People's Republic of China\\
$ ^3$Center for Gravitation and Cosmology, College of Physical Science and Technology,
Yangzhou University, Yangzhou 225009, China}

\begin{abstract}
\baselineskip=0.4 cm

{\bf Abstract} We give a brief review on the formation and the calculation of black hole shadow. Firstly, we introduce the conception of black hole shadow and the current works on a variety of black hole shadows. Secondly, we present main methods of calculating photon sphere radius and shadow radius, and then explain how the photon sphere affect the boundary of black hole shadow. We review the analytical calculation for black hole shadows which have analytic expressions for shadow boundary due to the integrable photon motion system. And we introduce the fundamental photon orbits which can explain the patterns of black hole shadow shape. Finally, we review the numerical calculation of black hole shadows with the backward ray-tracing method, and introduce some chaotic black hole shadows with self-similar fractal structures. Since the gravitational waves from the merger of binary black holes have been detected, we introduce a couple of shadows of binary black holes, which all have the eyebrowlike shadows around the main shadows with the fractal structures. We discuss the invariant phase space structures of photon motion system in black hole space-time, and explain the formation of black hole shadow is dominated by the invariant manifolds of certain Lyapunov orbits near the fixed points.

\end{abstract}

 \maketitle
\newpage
\section{Introduction}

In 2019, Event Horizon Telescope (EHT) Collaboration captured the first image of the supermassive black hole in the center of the giant elliptical galaxy M87 \cite{eht,fbhs1,fbhs2,fbhs3,fbhs4,fbhs5,fbhs6}, which opens a new era in the fields of astrophysics and black hole physics. Nowadays, more and more researchers have devoted themselves to the research of black hole shadows. The dark region in the center of black hole image is black hole shadow. The dark shadow appears because the light rays close to black hole are captured by black hole, thereby leaving a black shadow in the observer's sky. The trajectory of light propagation is determined by the structure of black hole space-time, so black hole shadow will carry much information of black hole space-time. For example, it is a perfect black disk for Schwarzschild black hole shadow; it gradually becomes a ``D"-shaped silhouette with the increase of spin parameter for Kerr black hole shadow\cite{sha2,sha3}; it is a cusp shadow for a Kerr black hole with Proca hair \cite{fpos2} and a Konoplya-Zhidenko rotating non-Kerr black hole\cite{sb10}. The research of black hole shadows plays a vital role in the study of black holes (constraining black hole parameters)\cite{sha8,sha9}, probing some fundamental physics issues including dark matter\cite{drk, polar7, polar8} and verification of various gravity theories\cite{lf,sha10,fR, 2101, 2107, 2111}.

Nowadays, a variety of black hole shadows have been investigated in Refs.\cite{sha2,sha3,fpos2,sb10,sha8,sha9,drk,polar7,polar8,lf,sha10,fR,2101,2107,2111,sw,swo,astro,chaotic,zhengwen23,sha18,binary,bsk,my,sMN,swo7,mbw,mgw,schm,scc,sha4,sha5,sha6,sha7,sha11,sha111,sha12,sha13,sha14,who,whr,sha141,sha15,sha16,sb1,sha17,sha19,sha191,sha192,sha193,sha194,shan1,shan1add,shan2add,shan3add,rr,pe,lf2,Zeng2020vsj,Zeng2020dco,lens,knn,bieu,ssp,lumi,nake,nakeop,BI}. The shadows of Schwarzschild and Kerr black hole immersed in Melvin magnetic field become more elongated in the horizontal direction as the magnetic field parameter increases\cite{schm,scc}. Some mediums or something else in universe could affect the light propagation and thus affect black hole shadow. V. Perlick\cite{sha16} took a non-magnetized pressure-less electron-ion plasma into account to research the influence of the plasma on the shadow of a spherically symmetric black hole. S. Chen\cite{pe} found the polarization of light in a special bumblebee vector field, and researched Kerr black hole shadow casted by the polarized lights. For black hole shadow in an expanding universe, A. Grenzebach\cite{knn} researched shadows of Kerr-Newman-NUT black holes with a cosmological constant, V. Perlick\cite{bieu} researched the shadow in the Kottler (Schwarzschild-de Sitter) space-time for comoving observers, P. C. Li\cite{ssp} researched the shadow of a spinning black hole in an expanding universe. The first image of a Schwarzschild black hole surrounded by a shining and rotating accretion disk was calculated by Luminet\cite{lumi}. We consider a solution of the superposition of a Schwarzschild black hole with Bach-Weyl ring, and studied the shadows of Schwarzschild black hole with Bach-Weyl ring\cite{mbw}. P. V. P. Cunha \cite{lens} researched the shadows of a black hole surrounded by a heavy Lemos-Letelier accretion disk. L. Amarilla\cite{sha10}, S. A. Dastan\cite{fR} and L. Fen\cite{lf} have researched the shadow of a rotating black hole in extended Chern-Simons modified gravity, in $f(R)$ gravity, and in quadratic Degenerate Higher Order Scalar Tensor (DHOST) theory respectively. We also studied the effect of a special polar gravitational wave on shadow of a Schwarzschild black hole\cite{mgw}. Furthermore, the shadows of wormhole and naked singularity also be investigated in Ref.\cite{sha14, who, whr} and Ref.\cite{nake, nakeop} respectively. It is hoped that these information imprinted in black hole shadow can be captured in the future astronomical observations including the upgraded Event Horizon Telescope and BlackHoleCam\cite{bhc} to study black holes and verify various gravity theories.

The black hole shadows investigated in many works are calculated analytically since they have analytic expressions for boundary due to the integrable photon motion system. Taking Kerr black hole shadow for instance, it has a third motion constant, the Carter constant\cite{carter}, to make the number of integration constants is equal to the degrees of freedom of the photon motion system. It is very convenient to study the black hole shadows. Oppositely, the black hole shadows can only be calculated numerically by the backward ray-tracing method\cite{sw,swo,astro,chaotic,my,sMN,lf,swo7,mbw,mgw,scc} for the photon motion system with non-integrable. Some null geodesic motions can even be chaotic, thus the black hole shadow appears chaotic phenomenon. The self-similar fractal structures appear in the black hole shadow originating from the chaotic lensing for a rotating black hole with scalar hair \cite{sw,swo,astro,chaotic}, Bonnor black diholes with magnetic dipole moment \cite{my}, a non-Kerr rotating compact object with quadrupole mass moment \cite{sMN} and binary black hole system \cite{zhengwen23,sha18,binary,bsk}.

The structure of this short review is as follows. In section 2, we briefly introduce the formation of black hole shadow, and take Schwarzschild black hole shadow as an example to explain that the photon sphere determines the boundary of shadow. In Section 3, we take Kerr black hole shadow as an example to review the analytical calculation for black hole shadows, and introduce the fundamental photon orbits which can explain the patterns of black hole shadow shape. In Section 4, we review the numerical calculation of black hole shadows with the backward ray-tracing method, and introduce some chaotic black hole shadows casted by chaotic lensing. We also introduce a couple of shadows of binary black holes and the relationship between the invariant phase space structures of photon motion system and black hole shadow. Our summary are given in Section 5.

\section{The boundary of black hole shadow: the photon sphere}

\begin{figure}
\subfigure[]{\includegraphics[width=4.1cm ]{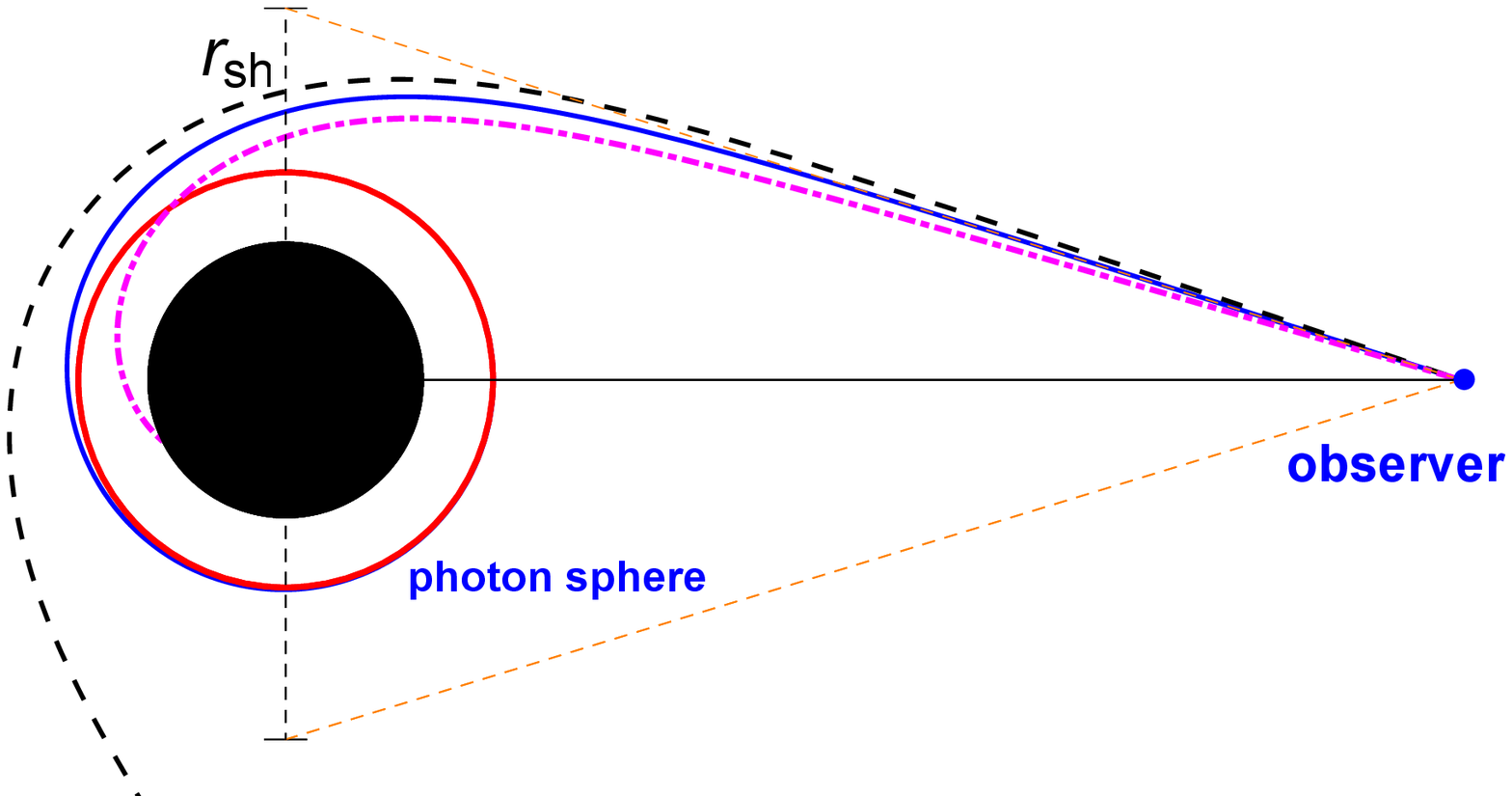}}\subfigure[]{\includegraphics[width=3.9cm ]{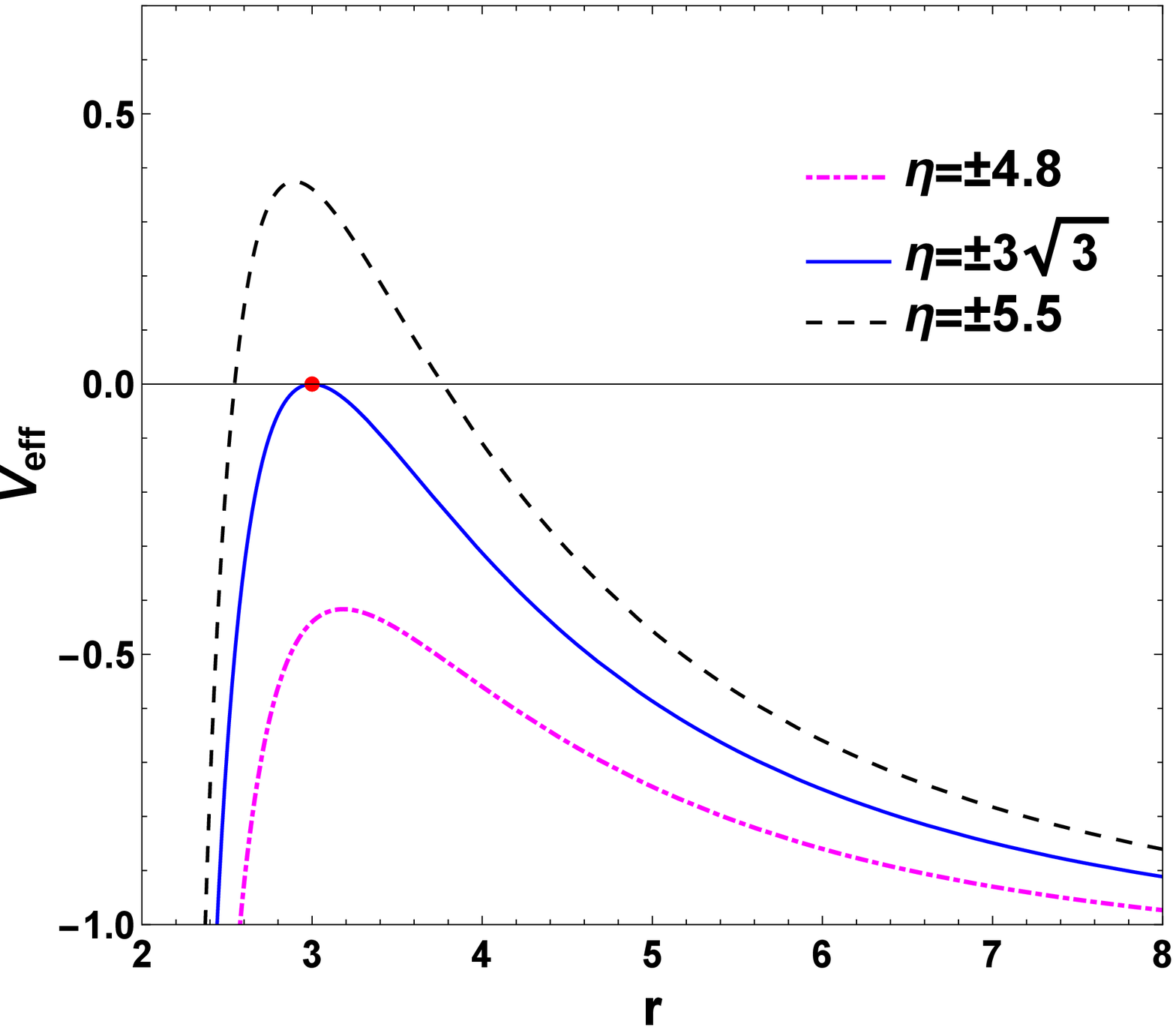}}
\caption{(a)The black disk represents Schwarzschild black hole, the red circle represents the photon sphere $r_{\rm ps}=3M$, the black dotted line represents black hole shadow with a radius $r_{\rm sh}=3\sqrt{3}M$.(b)The plot of the effective potential $V_{\rm eff}$ with the radial coordinate $r$ for different impact parameter $\eta$ in Schwarzschild black hole space-time. The red dot represents the photon sphere with $r_{\rm ps}=3M$.}
\label{pst}
\end{figure}

Black hole shadow is a dark silhouette observed in the sky, which corresponds to the light rays are captured by black hole. There may be a misconception that the boundary of black hole, event horizon, determines the boundary of black hole shadow. In point of fact, black hole shadow is about two and a half times larger than the event horizon in angular size\cite{eht,fbhs1}. There are two reasons for this situation. Firstly, the boundary of shadow is determined by the photon sphere outside the event horizon, which is compose of unstable photon circular orbits(the photon sphere radius $r_{\rm ps}=3M$ in the Schwarzschild case, $M$ is black hole mass). Secondly, the bending of light will make the photon sphere look bigger for observer, which are shown in Fig.\ref{pst}(a). In this figure, the black disk represents Schwarzschild black hole, the red circle represents the photon sphere. We assume light emitted from the observer backward in time. The light rays (magenta dotted line) that enter the photon sphere will be captured by black hole; the light rays (black dash line) that do not enter the photon sphere will fly away to infinity; the light rays (blue line) that spiral asymptotically towards the photon sphere will determine the boundary of black hole shadow. For the observer, they will see the black hole shadow with a radius $r_{\rm sh}=3\sqrt{3}M$ (see the black dash line in Fig.\ref{pst}(a)) for the Schwarzschild black hole.

Now, we take the Schwarzschild black hole as an example to calculate the photon sphere radius $r_{\rm ps}$ and shadow radius $r_{\rm sh}$. The Schwarzschild metric reads as
\begin{eqnarray}
\label{swdg}
{\rm d}s^{2}=-A(r){\rm d}t^{2}+B(r){\rm d}r^{2}+C(r)({\rm d}\theta^{2}+\sin^{2}\theta {\rm d}\varphi^{2}),
\end{eqnarray}
where $A(r)=1-\frac{2M}{r}, B(r)=\frac{1}{1-2M/r}, C(r)=r^{2}$. The Hamiltonian $\mathcal{H}$ of a photon propagation along null geodesics in Schwarzschild black hole space-time (\ref{swdg}) can be expressed as
\begin{eqnarray}
\label{hmd}
\mathcal{H}=\frac{1}{2}g^{\mu\nu}p_{\mu}p_{\nu}=\frac{1}{2}\bigg(\frac{p_{r}^{2}}{B(r)}+\frac{p_{\theta}^{2}}{C(r)}+V_{\rm eff}\bigg)=0,
\end{eqnarray}
where the effective potential $V_{\rm eff}$ is defined as
\begin{eqnarray}
\label{veff}
V_{\rm eff}&=&-\frac{E^{2}}{A(r)}+\frac{L_{z}^{2}}{C(r)\sin^{2}\theta}\nonumber\\
&=&\frac{E^{2}}{A(r)C(r)\sin^{2}\theta}\bigg(A(r)\eta^{2}-C(r)\sin^{2}\theta\bigg).\nonumber\\
\end{eqnarray}
$E=-p_{t}=-A(r)\dot{t}$ and $L_{z}=p_{\phi}=C(r)\sin^{2}\theta\dot{\varphi}$ are two constants of motion for the null geodesics motion, i.e., the energy and the $z$-component of the angular momentum, so the impact parameter $\eta=L_{z}/E$ also is a constant for the photon motion.

The spherical photon orbits on photon sphere must satisfy
\begin{eqnarray}
\label{ps}
V_{\rm eff}=0,\;\;\;\;\;\;\;\;\frac{\partial V_{\rm eff}}{\partial r}=0.
\end{eqnarray}
Moreover, the spherical photon orbits with $\partial^{2}V_{\rm eff}/\partial r^{2}<0$ are unstable; the spherical photon orbits with $\partial^{2}V_{\rm eff}/\partial r^{2}>0$ are stable. Without loss of generality, we consider the spherical photon orbits on the equatorial plane also known as light rings ($\theta=\pi/2$). Solving the equations (\ref{ps}), one can get the photon sphere radius $r_{\rm ps}=3M$ and the impact parameter of the spherical photon orbits on photon sphere $\eta_{\rm ps}=\pm3\sqrt{3}M$. The impact parameter $\eta>0$ represents the photon rotates counterclockwise for the observer in the north pole of Schwarzschild black hole; the $\eta<0$ represents the photon rotates clockwise. Fig.\ref{pst}(b) shows the plot of the effective potential $V_{\rm eff}$ with the radial coordinate $r$ for different impact parameter $\eta$ in Schwarzschild black hole space-time. The blue line in this figure represents the curve of $V_{\rm eff}$ with $\eta=\eta_{\rm ps}=\pm3\sqrt{3}M$, in which there is a point for $V_{\rm eff}=\frac{\partial V_{\rm eff}}{\partial r}=0$ and $\partial^{2}V_{\rm eff}/\partial r^{2}<0$ representing the unstable spherical photon orbits (photon sphere, $r=r_{\rm ps}=3M$), marked by red dot. The Hamiltonian $\mathcal{H}$ (\ref{hmd}) implies that the region of the effective potential $V_{\rm eff}>0$ is forbidden region for the photon motion. From Fig.\ref{pst}(b), one can infer that the photons with impact parameter $|\eta|>|\eta_{\rm ps}|$ can't reach the event horizon from outside the photon sphere, only the photons with $|\eta|<|\eta_{\rm ps}|$ can reach the event horizon, it also illustrates that the photon sphere determines the boundary of black hole shadow.

\begin{figure}
\includegraphics[width=8cm ]{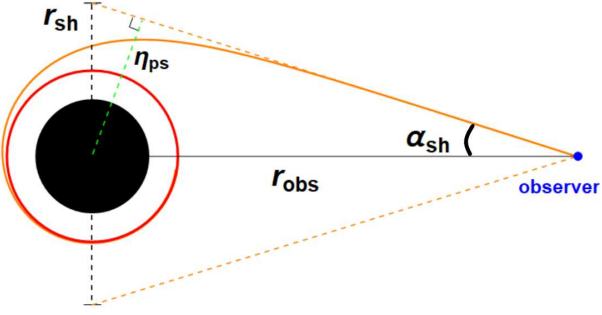}
\caption{The angular radius $\alpha_{\rm sh}$ of black hole shadow and the impact parameter $\eta_{\rm ps}$ for the light rays spiraling asymptotically towards the photon sphere.}
\label{pst2}
\end{figure}

In Fig.\ref{pst2}, we show the angular radius $\alpha_{\rm sh}$ of shadow for Schwarzschild black hole, which is calculated by Synge\cite{synge} as
\begin{eqnarray}
\label{ash}
\sin\alpha_{\rm sh}=\frac{3\sqrt{3}M\sqrt{1-2M/r_{\rm obs}}}{r_{\rm obs}}\;\;\;\; \text{for} \;\; r_{\rm obs}\geq3M.
\end{eqnarray}
We also label the impact parameter $\eta_{\rm ps}$ for the light rays spiraling asymptotically towards the photon sphere in Fig.\ref{pst2}, the green dotted line.
For the distant observer, the angular radius $\alpha_{\rm sh}$ (\ref{ash}) of black hole shadow can be rewritten as
\begin{eqnarray}
\label{ashy}
\sin\alpha_{\rm sh}=\frac{|\eta_{\rm ps}|}{r_{\rm obs}}\approx\alpha_{\rm sh}\;\;\;\; \text{for}  \;\; r_{\rm obs}\gg M,
\end{eqnarray}
and
\begin{eqnarray}
\label{ashy1}
\tan\alpha_{\rm sh}=\frac{r_{\rm sh}}{r_{\rm obs}}\approx\alpha_{\rm sh}\;\;\;\; \text{for} \;\; r_{\rm obs}\gg M.
\end{eqnarray}
So the impact parameter of the photon sphere is approximately equal to the radius of the black hole shadow, $r_{\rm sh}\approx|\eta_{\rm ps}|$. It also could be derived out by moving the observer to infinity in Fig.\ref{pst2}.

\section{Analytical calculation for black hole shadows}
Many of the current works on black hole shadows have only investigated the case that the geodesic equations for photons can separate variables, and the number of integration constants is equal to the degrees of freedom of the photon motion system, hence the system is integrable. Such as the photon motion system in Kerr black hole space-time has a third motion constant, the Carter constant\cite{carter}, except for the energy $E$ and the $z$-component of the angular momentum $L_{z}$. Consequently, black hole shadows in the integrable photon motion systems have analytic expressions for the boundary, and make it very convenient to study the shadows of black holes.

\subsection{Shadow of a Kerr black hole}

Now, we take Kerr black hole for example to analytically calculate black hole shadow. In the Boyer-Lindquist coordinates, the Kerr metric has a form
\begin{eqnarray}
\label{kerr}
{\rm d}s^{2}&&=-(1-\frac{2Mr}{\rho^{2}}){\rm d}t^{2}+\frac{\rho^{2}}{\Delta}{\rm d}r^{2}+\rho^{2} {\rm d}\theta^{2}\\ \nonumber &&+\sin^{2}\theta\bigg(r^{2}+a^{2}+\frac{2Mra\sin^{2}\theta}{\rho^{2}}\bigg){\rm d}\phi^{2}\\ \nonumber
&&-\frac{4Mra\sin^{2}\theta}{\rho^{2}}{\rm d}t{\rm d}\phi,
\end{eqnarray}
where
\begin{equation}
\Delta=a^{2}+r^{2}-2Mr,\;\;\;\;\;\;\;\;\;\; \rho^{2}=r^{2}+a^{2}\cos^{2}\theta.
\end{equation}
$M$ and $a$ is the mass and spin parameter of Kerr black hole. The Hamiltonian of a photon propagation along null geodesics in a Kerr black hole space-time can be expressed as
\begin{eqnarray}
\label{hami}
H(x,p)&=&\frac{1}{2}g^{\mu\nu}(x)p_{\mu}p_{\nu}=\frac{\Delta}{2\rho^{2}}P_{r}^{2} \nonumber\\
&+& \frac{1}{2\Delta\rho^{2}}[(r^{2}+a^{2})P_{t}+aP_{\phi}]^{2}+\frac{1}{2\rho^{2}}P_{\theta}^{2}\nonumber\\ &+&\frac{1}{2\rho^{2}\sin^{2}\theta}
(P_{\phi}+aP_{t}\sin^{2}\theta)^{2}=0.
\end{eqnarray}
The two conserved quantities in photon motion: the energy $E$ and the z-component of the angular momentum $L_{z}$ can be obtained as
\begin{eqnarray}
\label{EL}
E&=&-p_{t}=-g_{tt}\dot{t}-g_{t\phi}\dot{\phi},\nonumber\\
L_{z}&=&p_{\phi}=g_{\phi\phi}\dot{\phi}+g_{\phi t}\dot{t}.
\end{eqnarray}
With these two conserved quantities, the null geodesic equations can be written as\cite{carter}
\begin{eqnarray}
&&\dot{t}=E+\frac{2Mr(a^{2}E-aL_{z}+Er^{2})}{\Delta\rho^{2}},\nonumber \\
&&\dot{\phi}=\frac{2MraE}{\Delta\rho^{2}}+\frac{\Delta-a^{2}\sin^{2}\theta}{\Delta\rho^{2}\sin^{2}\theta}L_{z},\nonumber \\
&&R(r)=\rho^{4}\dot{r}^{2}=\Delta^{2} p_{r}^{2}=-\Delta[Q+(aE-L_{z})^{2}]\nonumber \\
&&\quad\quad \quad+[aL_{z}-(r^{2}+a^{2})E]^{2},\nonumber \\
&&\Theta(\theta)=\rho^{4}\dot{\theta}^{2}=p_{\theta}^{2}=Q-\cos^{2}\theta
\bigg(\frac{L_{z}^{2}}{\sin^{2}\theta}-a^{2}E^{2}\bigg).\nonumber \\
\label{tfc4}
\end{eqnarray}
where the quantity $Q$ is the Carter constant that is the third conserved quantity in the null geodesics in Kerr black hole space-time\cite{carter}. It is the Carter constant $Q$ makes it possible to separate the variables in the null geodesic equations (\ref{tfc4}), and the photon motion system is integrable. The photon sphere determining the boundary of black hole shadow satisfies
\begin{eqnarray}
\label{rr1}
\dot{r}=0, \;\;\;\;\;\text{and} \;\;\;\;\;\ddot{r}=0,
\end{eqnarray}
which yields
\begin{eqnarray}
\label{R}
R(r)&=&-\Delta[Q+(aE-L_{z})^{2}] \nonumber\\&&
+[aL_{z}-(r^{2}+a^{2})E]^{2}=0, \nonumber\\
R'(r)&=&-4Er[aL_{z}-(r^{2}+a^{2})E]\nonumber\\ &&-2 (r-M)[Q+(aE-L_{z})^{2}]=0.
\end{eqnarray}
Solving the equations (\ref{R}), we find that for the spherical orbits motion of photon the impact parameter $\eta$ and the constant $\sigma=Q/E^{2}$ have the form
\begin{eqnarray}
\eta&=&\frac{L_{z}}{E}=-\frac{r^{2}(r-3M)+a^{2}(r+M)}{a(r-M)}, \nonumber \\
\sigma&=&\frac{Q}{E^{2}}=\frac{r^{3}[4a^{2}M-r(r-3M)^{2}]}{a^{2}(r-M)^{2}}.
\label{cs}
\end{eqnarray}

\begin{figure}
\center{\includegraphics[width=7cm ]{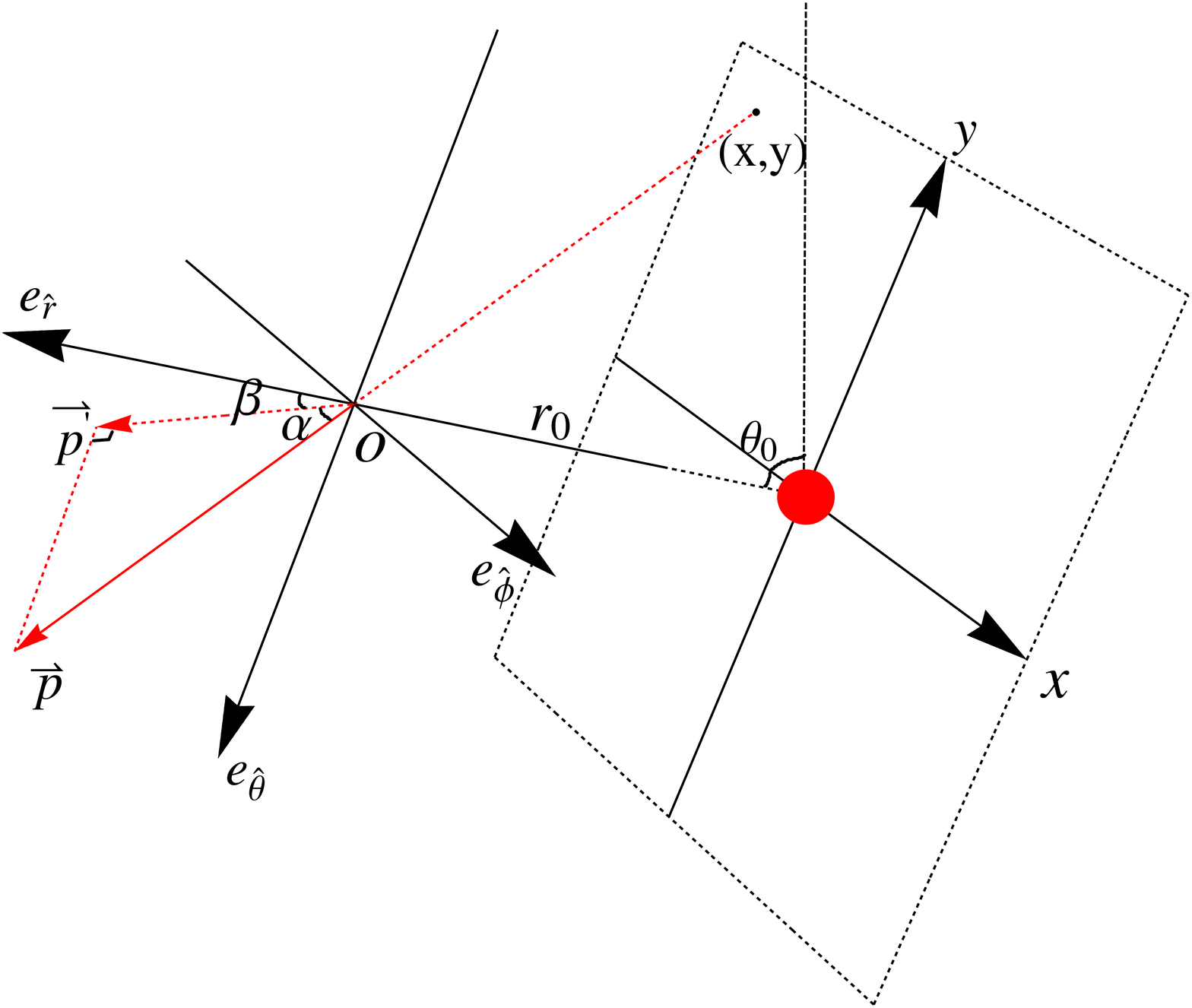}
\caption{Perspective drawing of the geometric projection of the photon's linear momentum $\vec{p}$ in
the observer's frame $\{e_{\hat{r}},e_{\hat{\theta}},e_{\hat{\phi}}\}$. The red sphere and the point $ O (r_{0},\theta_{0},0)$ denote
the position of  black hole and observer, respectively. The vector $\vec{p'}$ is the projection of $\vec{p}$ onto plane $(e_{\hat{r}},e_{\hat{\phi}})$ and  $\alpha$ is the angle between $\vec{p}$ and plane $(e_{\hat{r}},e_{\hat{\phi}})$, $\beta$ is the angle between $\vec{p'}$ and basis $e_{\hat{r}}$.}
\label{zb}}
\end{figure}

\begin{figure}
\subfigure[]{\includegraphics[width=4cm ]{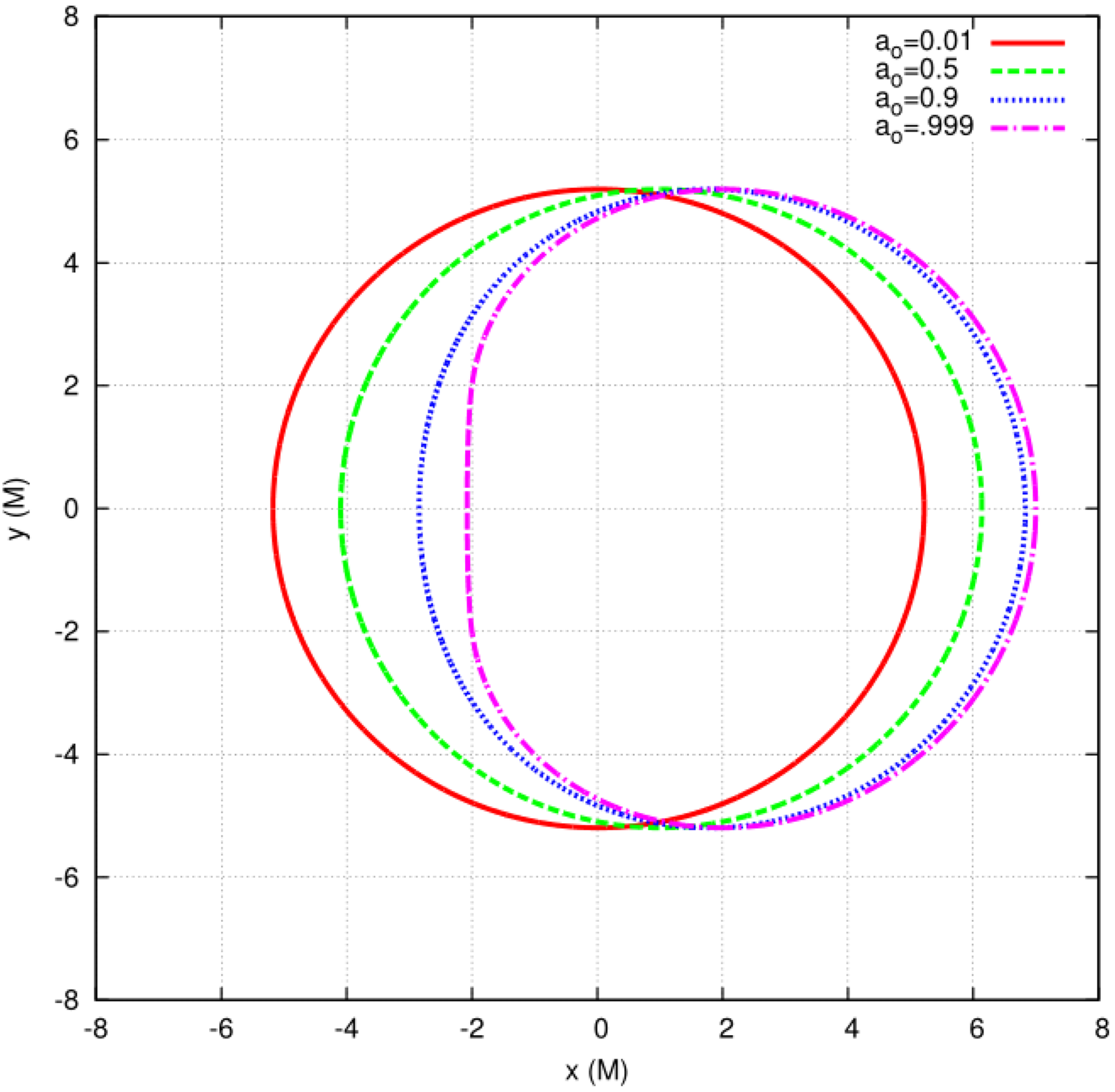}}\subfigure[]{\includegraphics[width=4.0cm ]{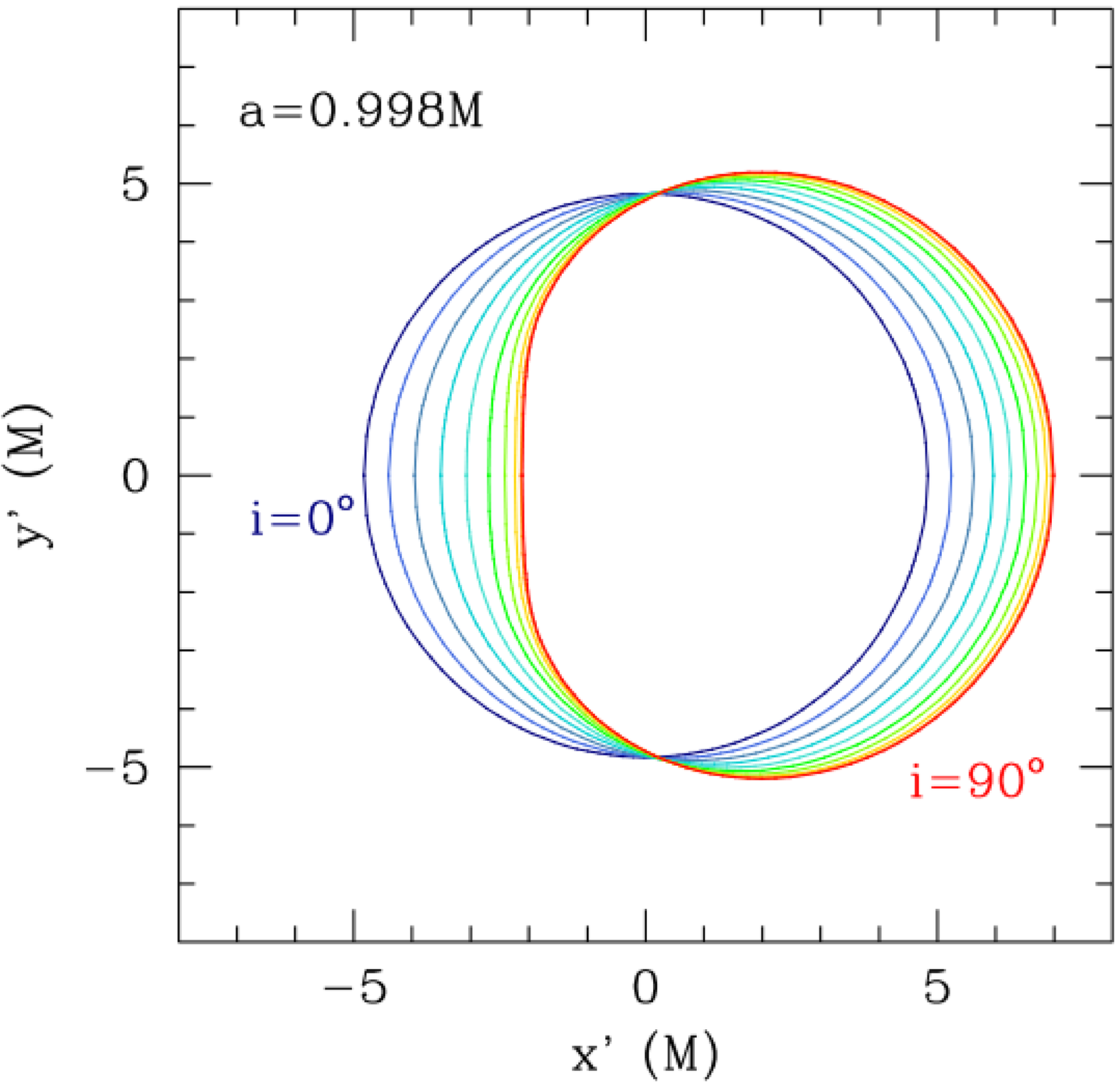}}
\caption{(a) The shadows of Kerr black hole with different spin parameter $a$ for the observer's inclination angle $i$ ($\theta_{0}$)=$\pi/2$\cite{swo}. (b)The shadows of Kerr black hole with different observer's inclination angle $i$ for $a=0.998M$\cite{swo7}.}
\label{kerrt}
\end{figure}

We assume that the static observer is locally at ($r_{o}, \theta_{o}$) in zero-angular-moment-observers (ZAMOs) reference frame \cite{sha2}, in which the observer can determine the coordinates of the photons in the sky. The observer basis $\{e_{\hat{t}},e_{\hat{r}},e_{\hat{\theta}},e_{\hat{\phi}}\}$ can be expanded in the coordinate basis $\{ \partial_t,\partial_r,\partial_{\theta},\partial_{\phi} \}$ as a form \cite{sw,swo,astro,chaotic,my,sMN,lf,swo7,mbw,mgw,scc}
\begin{eqnarray}
\label{zbbh}
e_{\hat{\mu}}=e^{\nu}_{\hat{\mu}} \partial_{\nu},
\end{eqnarray}
where the transform matrix $e^{\nu}_{\hat{\mu}}$ obeys to $g_{\mu\nu}e^{\mu}_{\hat{\alpha}}e^{\nu}_{\hat{\beta}}
=\eta_{\hat{\alpha}\hat{\beta}}$, and $\eta_{\hat{\alpha}\hat{\beta}}$ is the metric of Minkowski space-time. Generally, it is convenient to choice a decomposition connected to the reference frame in relation to spatial infinity, which is given by \cite{sw,swo,astro,chaotic,my,sMN,lf,swo7,mbw,mgw,scc}
\begin{eqnarray}
\label{zbbh1}
e^{\nu}_{\hat{\mu}}=\left(\begin{array}{cccc}
\zeta&0&0&\gamma\\
0&A^r&0&0\\
0&0&A^{\theta}&0\\
0&0&0&A^{\phi}
\end{array}\right),
\end{eqnarray}
where $\zeta$, $\gamma$, $A^r$, $A^{\theta}$, and $A^{\phi}$ are real coefficients.
From the Minkowski normalization
\begin{eqnarray}
e_{\hat{\mu}}e^{\hat{\nu}}=\delta_{\hat{\mu}}^{\hat{\nu}},
\end{eqnarray}
one can obtain
\begin{eqnarray}
\label{xs}
&&A^r=\frac{1}{\sqrt{g_{rr}}},\;\;\;\;\;\;\;\;\;
A^{\theta}=\frac{1}{\sqrt{g_{\theta\theta}}},\;\;\;\;\;\;\;\;
A^{\phi}=\frac{1}{\sqrt{g_{\phi\phi}}},\nonumber\\
&&\zeta=\sqrt{\frac{g_{\phi \phi}}{g_{t\phi}^{2}-g_{tt}g_{\phi \phi}}},\quad\gamma=-\frac{g_{t\phi}}{g_{\phi\phi}}\sqrt{\frac{g_{\phi \phi}}{g_{t\phi}^{2}-g_{tt}g_{\phi \phi}}}.\nonumber\\
\end{eqnarray}
Therefore, one can get the locally measured four-momentum $p^{\hat{\mu}}$ of a photon by the projection of its four-momentum $p^{\mu}$  onto $e_{\hat{\mu}}$,
\begin{eqnarray}
\label{dl}
p^{\hat{t}}=-p_{\hat{t}}=-e^{\nu}_{\hat{t}} p_{\nu},\;\;\;\;\;\;\;\;\;
\;\;\;\;p^{\hat{i}}=p_{\hat{i}}=e^{\nu}_{\hat{i}} p_{\nu}.
\end{eqnarray}
With the help of Eq.(\ref{xs}), the locally measured four-momentum $p^{\hat{\mu}}$ can be further written as
\begin{eqnarray}
\label{kmbh}
p^{\hat{t}}&=&\zeta E-\gamma L,\;\;\;\;\;\;\;\;\;\;\;\;\;p^{\hat{r}}=\frac{1}{\sqrt{g_{rr}}}p_{r} ,\nonumber\\
p^{\hat{\theta}}&=&\frac{1}{\sqrt{g_{\theta\theta}}}p_{\theta},
\;\;\;\;\;\;\;\;\;\;\;\;\;\;\;\;
p^{\hat{\phi}}=\frac{1}{\sqrt{g_{\phi\phi}}}L.
\end{eqnarray}
The spatial position of observer in the black hole space-time is set to ($r_{o},\theta_{o},0$) as shown  in Fig. \ref{zb}.  The $3-$vector $\vec{p}$  is the photon's linear momentum with components $p_{\hat{r}}$, $p_{\hat{\theta}}$ and $p_{\hat{\phi}}$
in the orthonormal basis $\{e_{\hat{r}},e_{\hat{\theta}},e_{\hat{\phi}}\}$,
\begin{eqnarray}
\vec{p}=p^{\hat{r}}e_{\hat{r}}+p^{\hat{\theta}}
e_{\hat{\theta}}+p^{\hat{\phi}}e_{\hat{\phi}}.
\end{eqnarray}
According to the geometry of the photon's detection, we have
\begin{eqnarray}
\label{fl}
p^{\hat{r}}&=&|\vec{p}|\cos\alpha\cos\beta, \nonumber\\
p^{\hat{\theta}}&=&|\vec{p}|\sin\alpha, \nonumber\\
p^{\hat{\phi}}&=&|\vec{p}|\cos\alpha\sin\beta.
\end{eqnarray}
Actually, the angular coordinates $(\alpha, \beta)$ of a point in the observer's local sky define the direction of the associated light ray. The coordinates $(x,y)$ of a point in the observer's local sky are related to its angular coordinates $(\alpha, \beta)$ by \cite{sw,swo,astro,chaotic,my,sMN,lf,swo7,mbw,mgw,scc}
\begin{eqnarray}
\label{xd}
&&x=-r\tan\beta|_{(r_{0},\theta_{0})}=-r\frac{p^{\hat{\phi}}}{p^{\hat{r}}}|_{(r_{0},\theta_{0})}, \nonumber\\
&&y=r\frac{\tan\alpha}{\cos\beta}|_{(r_{0},\theta_{0})}=r\frac{p^{\hat{\theta}}}{p^{\hat{r}}}|_{(r_{0},\theta_{0})}.
\end{eqnarray}
In general, the observer is far away from the black hole, so we can take the limit $r_{0}\rightarrow\infty$. According the equations (\ref{tfc4}), (\ref{cs}), (\ref{kmbh}) and (\ref{xd}), we can get the analytic expressions for the boundary of Kerr black hole shadow
\begin{eqnarray}
\label{xd1w}
x&=&-\frac{\eta}{\sin \theta_{0}}, \nonumber\\
y&=&\pm\sqrt{\sigma+a^{2}\cos^{2}\theta_{0}-\xi^{2}\cot^{2}\theta_{0}}.
\end{eqnarray}
Fig.\ref{kerrt}(a) shows the shadows of Kerr black hole with different spin parameter $a$ for the observer's inclination angle $i$ ($\theta_{0}$)=$\pi/2$\cite{swo}. Fig.\ref{kerrt}(b) shows Kerr black hole shadows with different observer's inclination angle $i$ for $a=0.998M$\cite{swo7}. One can find the shadow of Kerr black hole is related to both the spin parameter $a$ of black hole and the observer's inclination angle $i$. As the spin parameter $a$ increases, the shape of black hole shadow gradually changes from a circle to a ``D" shape for the observer on the equatorial plane. The appearance of the ``D" shaped shadow is due to the fact that the radiuses of unstable prograde and retrograde light
rings (photon circular orbits on the equatorial plane) are different, which can be expressed as
\begin{eqnarray}
\label{rps}
r_{\rm lr}=2M\bigg(1+\cos\bigg[\frac{2}{3}\arccos(\pm\frac{a}{M})\bigg]\bigg).
\end{eqnarray}
It represents the radius of prograde (retrograde) photon circular orbits when taking the ``-" (``+") sign.

\begin{figure}
\center{\includegraphics[width=6cm ]{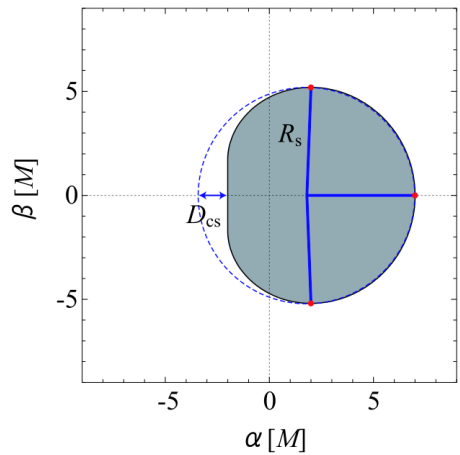}
\caption{The radius of Kerr black hole shadow $R_{\rm s}$ and the deviation $D_{\rm cs}$ of Kerr black hole shadow from a circle\cite{sha9}.}
\label{kerrcs}}
\end{figure}

\begin{figure}
\subfigure[$R_{s}$]{\includegraphics[width=4cm ]{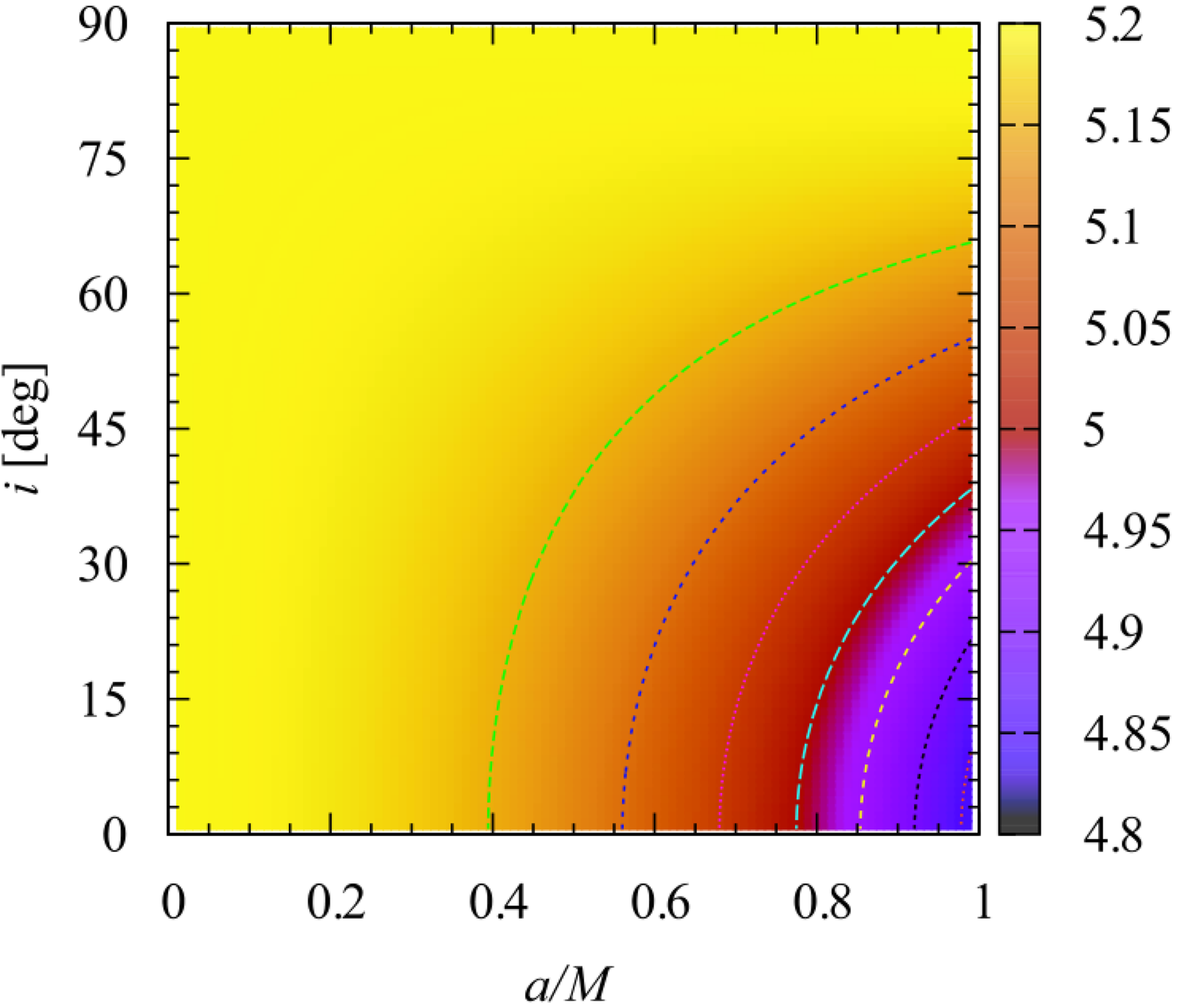}}\subfigure[$\delta_{\rm s}$]{\includegraphics[width=4cm ]{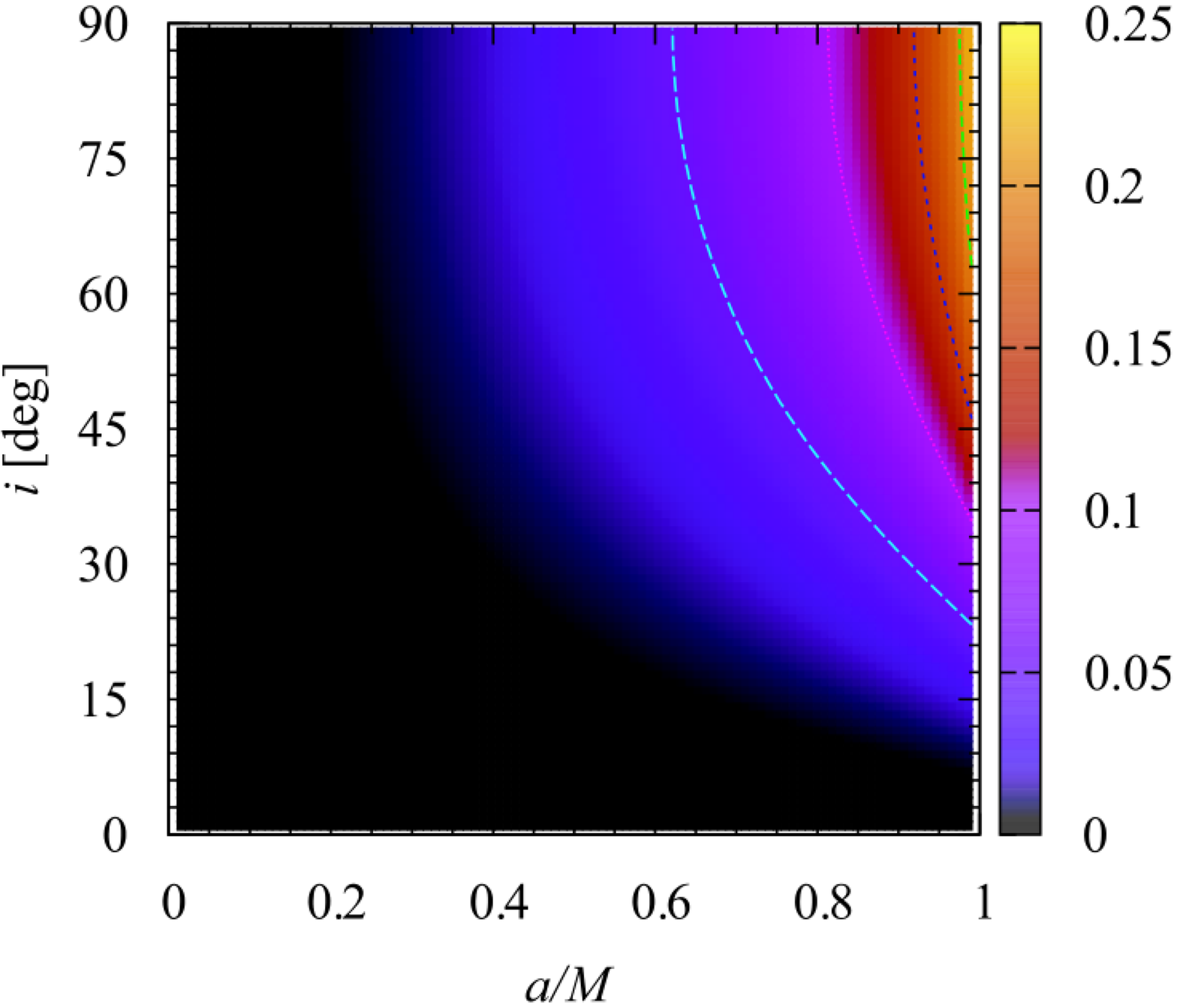}}
\caption{The contour plots of the radius $R_{\rm s}$ and deviation parameter $\delta_{\rm s}$ of Kerr black hole shadow as a function of the spin parameter $a$ and the observer's inclination angle $i$\cite{sha9}.}
\label{kerrrp}
\end{figure}

In addition, one can determine the spin parameter $a$ and the observer's inclination angle $\theta_{0}$ of Kerr black hole by calculating the deviation of Kerr black hole shadow from a circle. As shown in Fig.\ref{kerrcs}\cite{sha9}, $R_{\rm s}$ represents the radius of Kerr black hole shadow. The deviation parameter $\delta_{\rm s}=D_{\rm cs}/R_{\rm s}$ can represents the deviation of Kerr black hole shadow from a circle. Fig.\ref{kerrrp} shows the contour plots of the radius $R_{\rm s}$ and deviation parameter $\delta_{\rm s}$ of Kerr black hole shadow as a function of the spin parameter $a$ and the observer's inclination angle $i$\cite{sha9}. The plots indicate that $R_{\rm s}$ and $\delta_{\rm s}$ can be used as observable measurements to determine the parameters of black hole.

\subsection{Fundamental photon orbits}

\begin{figure}
 \subfigure[]{ \includegraphics[width=7.5cm ]{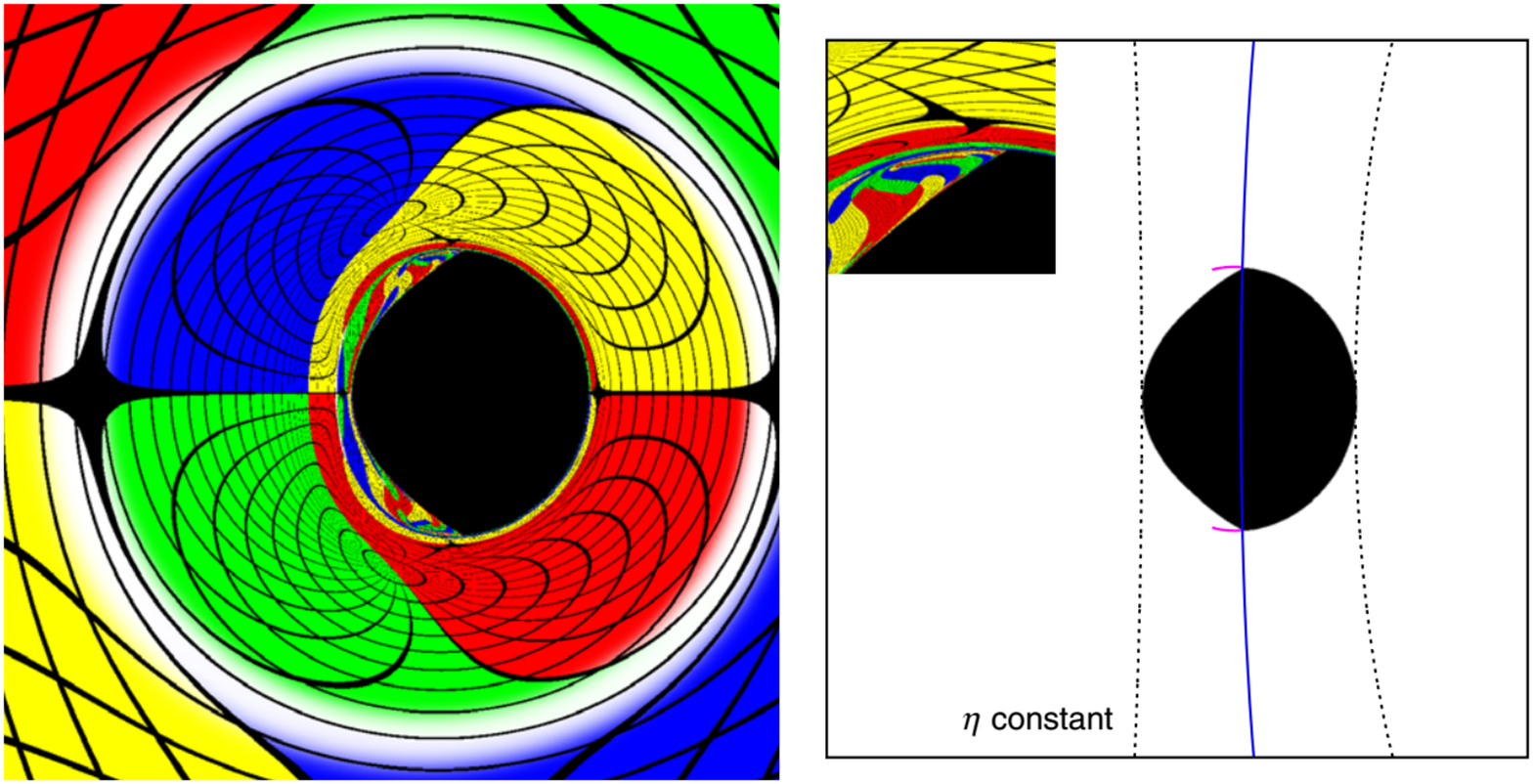}}\\
 \subfigure[]{ \includegraphics[width=7.5cm ]{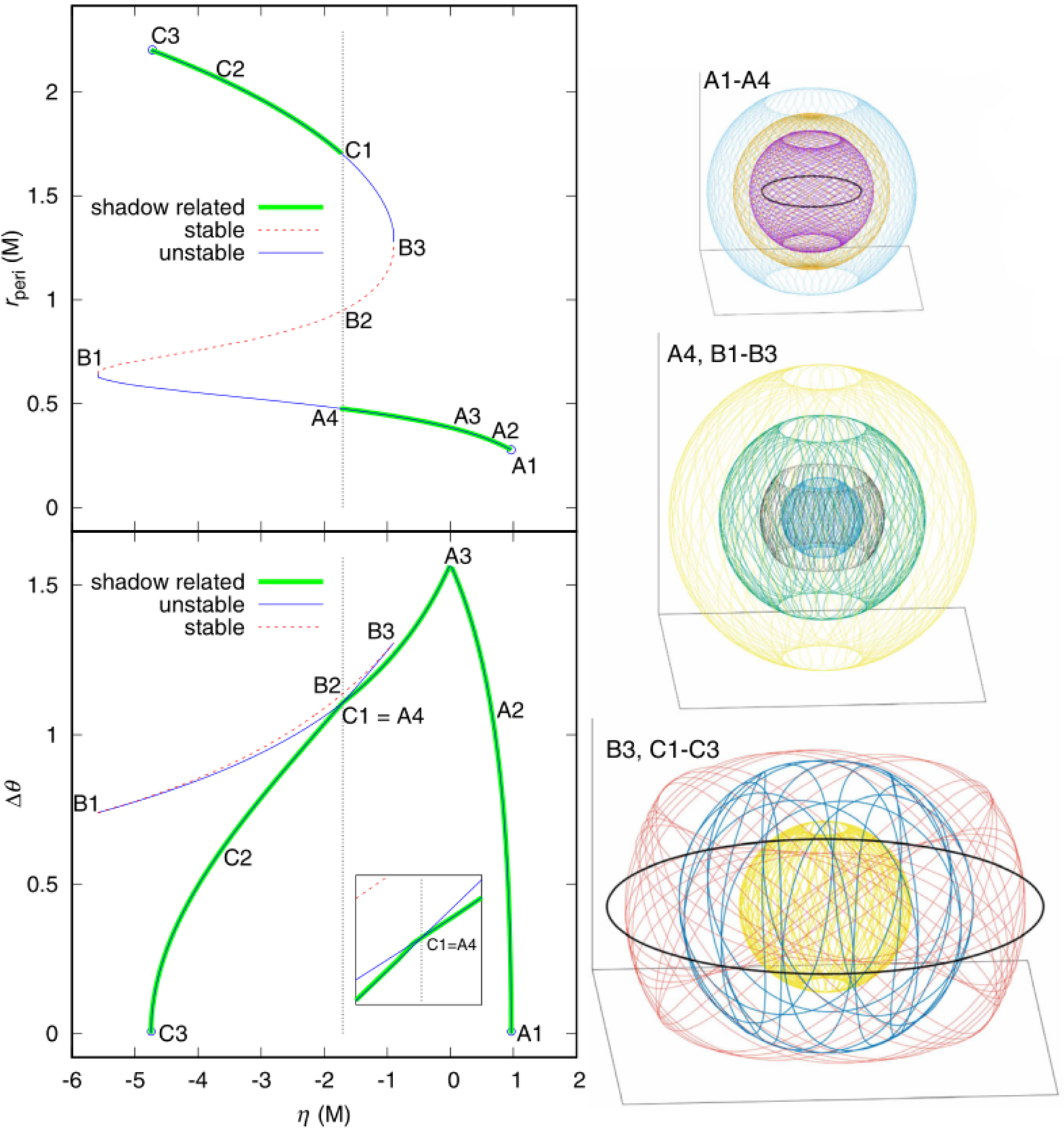}}
\caption{ The cusp shadow of a Kerr black hole with Proca hair and the fundamental photon orbits\cite{fpos2}.}
\label{sab}
\end{figure}

For a non-Kerr rotating black hole whose null geodesic equations is integrable, the black hole shadow can also be obtained in the same way for the calculation (\ref{hami}-\ref{xd1w}) of Kerr black hole shadow. P. V. P. Cunha et al\cite{fpos2} studied the shadow of a Kerr black hole with Proca hair\cite{fpos2}, and found black hole shadow has a cusp silhouette, shown in Fig.\ref{sab}(a). The further analysis shows that these novel patterns in shadow are related to the non-planar bound photon orbits in some generic stationary, axisymmetric space-time, namely the fundamental photon orbits (FPOs)\cite{fpos2}. Actually, a fundamental photon orbit is one of the photon circular orbits on the photon sphere; all the FPOs will make up the photon sphere determining the shadow boundary, so FPOs can explain the patterns of black hole shadow shape. The left panel of Fig.\ref{sab}(b)\cite{fpos2} shows $\Delta\theta\equiv|\theta_{\text{max}}-\frac{\pi}{2}|$ and $r_{\text{peri}}$ with the impact parameter $\eta$ of FPOs for the cusp shadow (Fig.\ref{sab}(a)), in which there are ten FPOs marked by ``A1-A4, B1-B3, C1-C3". Here $\theta_{\text{max}}$  denotes the maximal/minimal angular coordinate of a FPO, and $r_{\text{peri}}$ is the perimetral radius. The right panel of Fig.\ref{sab}(b) shows the spatial trajectories of these ten FPOs in Cartesian coordinates, which move around black hole. The FPOs A1 and C3 are the unstable prograde and retrograde light rings shown as two black circles on the equatorial plane. Other FPOs are non-planar bound photon orbits crossing the equatorial plane. The continuum of FPOs can be split into one stable branch (the red dotted line) and two unstable branches (the green and blue lines). Obviously, one can find  there exists a swallow-tail shape pattern related to FPOs in the $\eta-\Delta\theta$ plane, which yields a jump occurred at the FPOs A4 and C1. The discontinuity in these orbits (i.e., $r_{\text{peri}(C1)}>r_{\text{peri}(A4)}$) originating from this jump induces the emergence of the cusp shadow. It is shown that the unstable FPOs determine the boundary of shadow, and also can explain the patterns of shadow shape. We studied the shadows of a Konoplya-Zhidenko rotating non-Kerr black hole, also found the cusp edge for black hole shadow\cite{sb10}.
The fundamental photon orbits have been used to study black hole for general parameterized metrics\cite{parameter1,parameter2,parameter3,parameter4,parameter5,parameter6}, which could be beneficial to test Kerr hypothesis through black hole shadows.
 A plasma is a dispersive medium and the light rays deviate from lightlike geodesics in a way that depends on the frequency. The shadows of black holes in plasma have been also investigated in \cite{plasma1,plasma2,plasma3,plasma4,plasma5,plasma6,plasma7,plasma8,plasma9,plasma10,plasma11}.

\section{Numerical calculation for black hole shadows}
In fact, completely integrable systems are very few in the real world. Schwarzschild or Kerr black hole could be perturbed by some extra gravitational sources (accretion disks or rings), magnetic field, scalar hair and so on. According to the Kolmogorov-Arnold-Moser (KAM) theorem\cite{KAM}, if any integrable system gets perturbed, the system will no longer be integrable. Now, the black hole shadows can only be calculated numerically. It is expected that the non-integrable photon motions bring some new patterns and structures in black hole shadow.

\subsection{The backward ray-tracing method}
For the black holes whose photon motion system are non-integrable, black hole shadows can not be calculated analytically. It must make use of the backward ray-tracing method\cite{sw,swo,astro,chaotic,my,sMN,lf,swo7,mbw,mgw,scc} to obtain black hole shadow. In this method, the light rays are assumed to evolve from the observer backward in time and the information carried by each ray would be respectively assigned to a pixel in a final image in observer's sky. The light rays falling down into even horizon of black hole correspond to black pixels composing black hole shadow. With this spirit, we must solve numerically null geodesic equations
\begin{eqnarray}
\label{lgcdx}
\frac{{\rm d}^{2}x^{\alpha}}{{\rm d}\tau^{2}}+\Gamma^{\alpha}_{\mu\nu}\frac{{\rm d}x^{\mu}}{{\rm d}\tau}\frac{{\rm d}x^{\nu}}{{\rm d}\tau}=0,
\end{eqnarray}
where $\tau$ denotes the proper time.

\begin{figure}
\center{\includegraphics[width=6.5cm ]{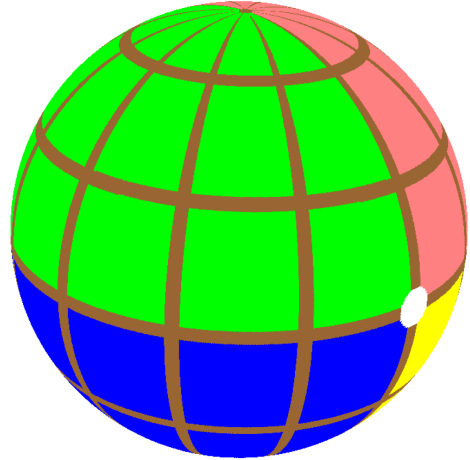}
\caption{The light-emitting celestial sphere marked by four different colored quadrants, the brown grids as longitude and latitude, and the white reference spot lies at the intersection of the four colored quadrants\cite{my,sha18,sw}.}
\label{gy}}
\end{figure}

One can set light-emitting celestial sphere as light source marked by four different colored quadrants, the brown grids as longitude and latitude, and the white reference spot lies at the intersection of the four colored quadrants, which is same as the celestial sphere in Ref.\cite{my,sha18,sw}, shown in Fig.\ref{gy}. Black hole is placed at the center of the light sphere. The observer is placed at the other intersection of the four colored quadrants.

\begin{figure}
\center{\subfigure[]{\includegraphics[width=4cm ]{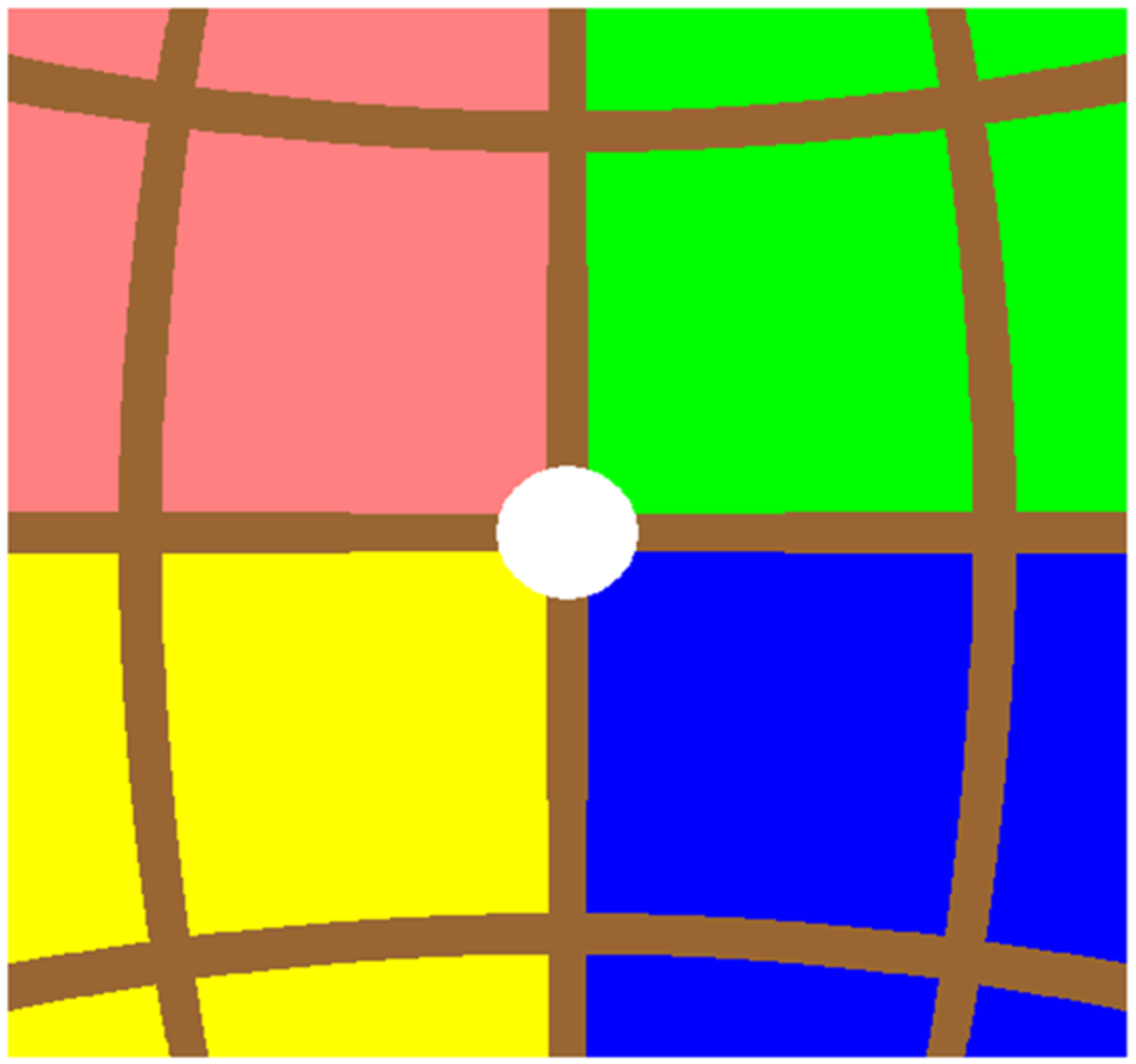}}\subfigure[]{\includegraphics[width=4cm ]{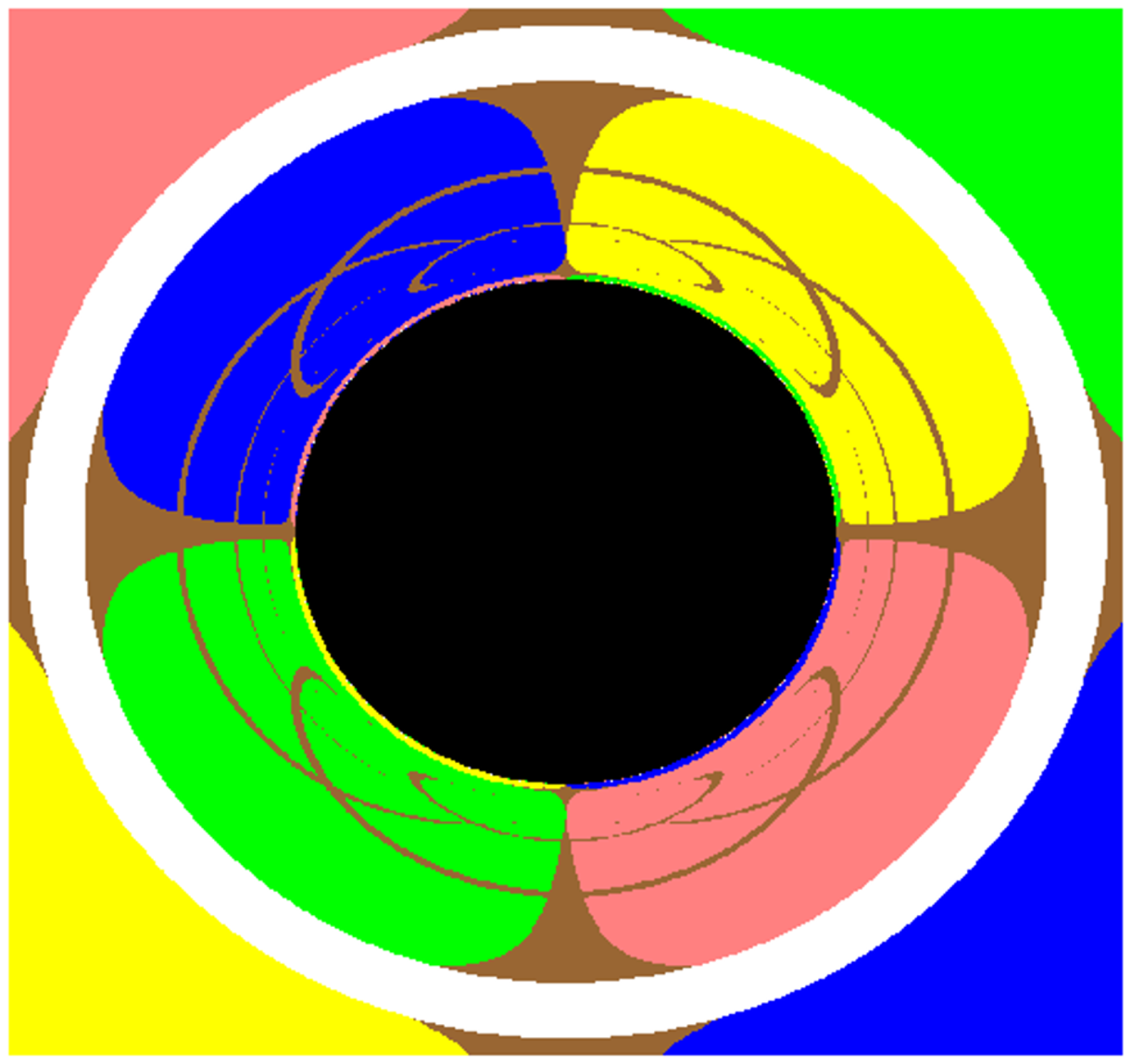}}\\
\subfigure[]{\includegraphics[width=4cm ]{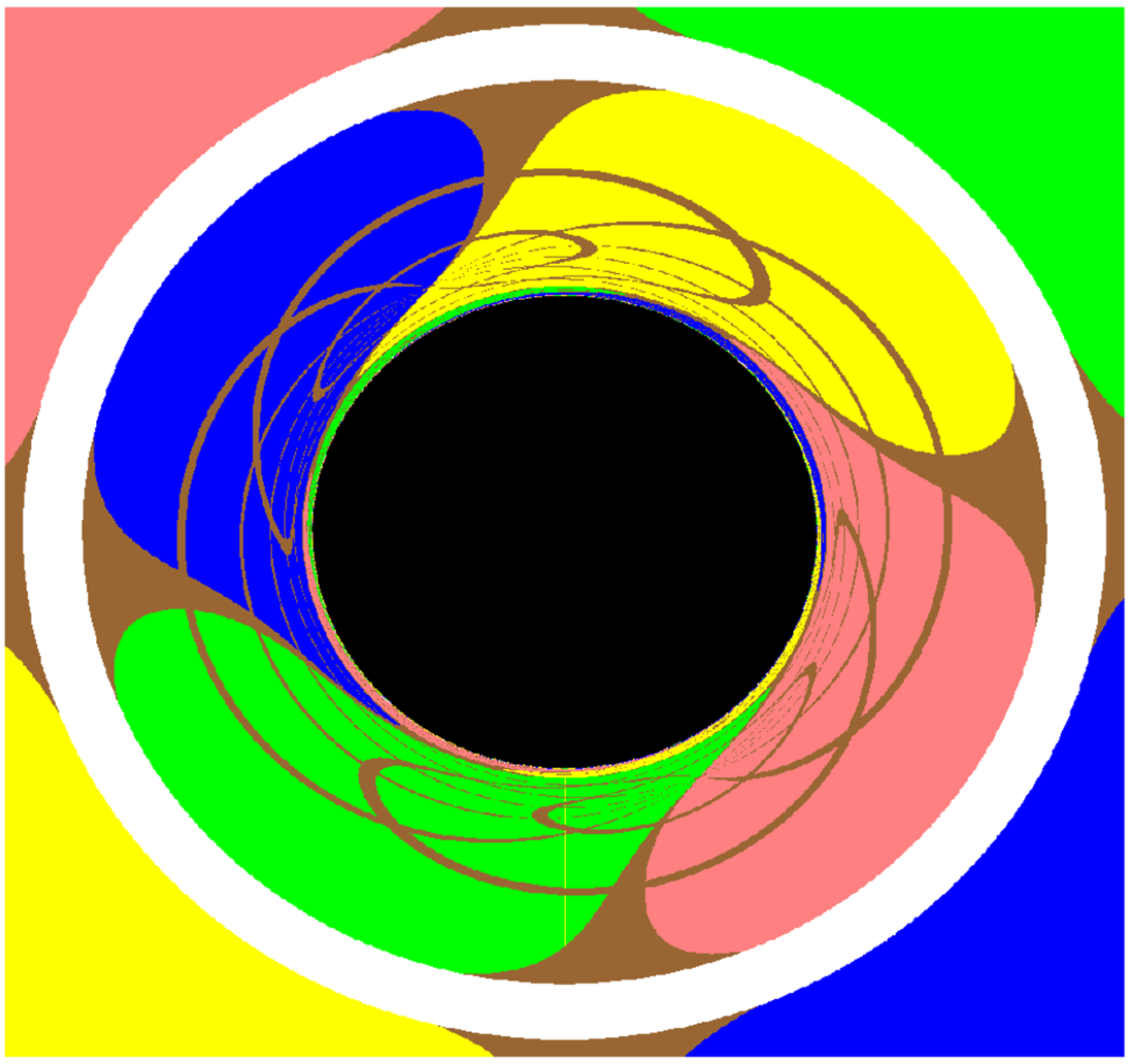}}\subfigure[]{\includegraphics[width=4cm ]{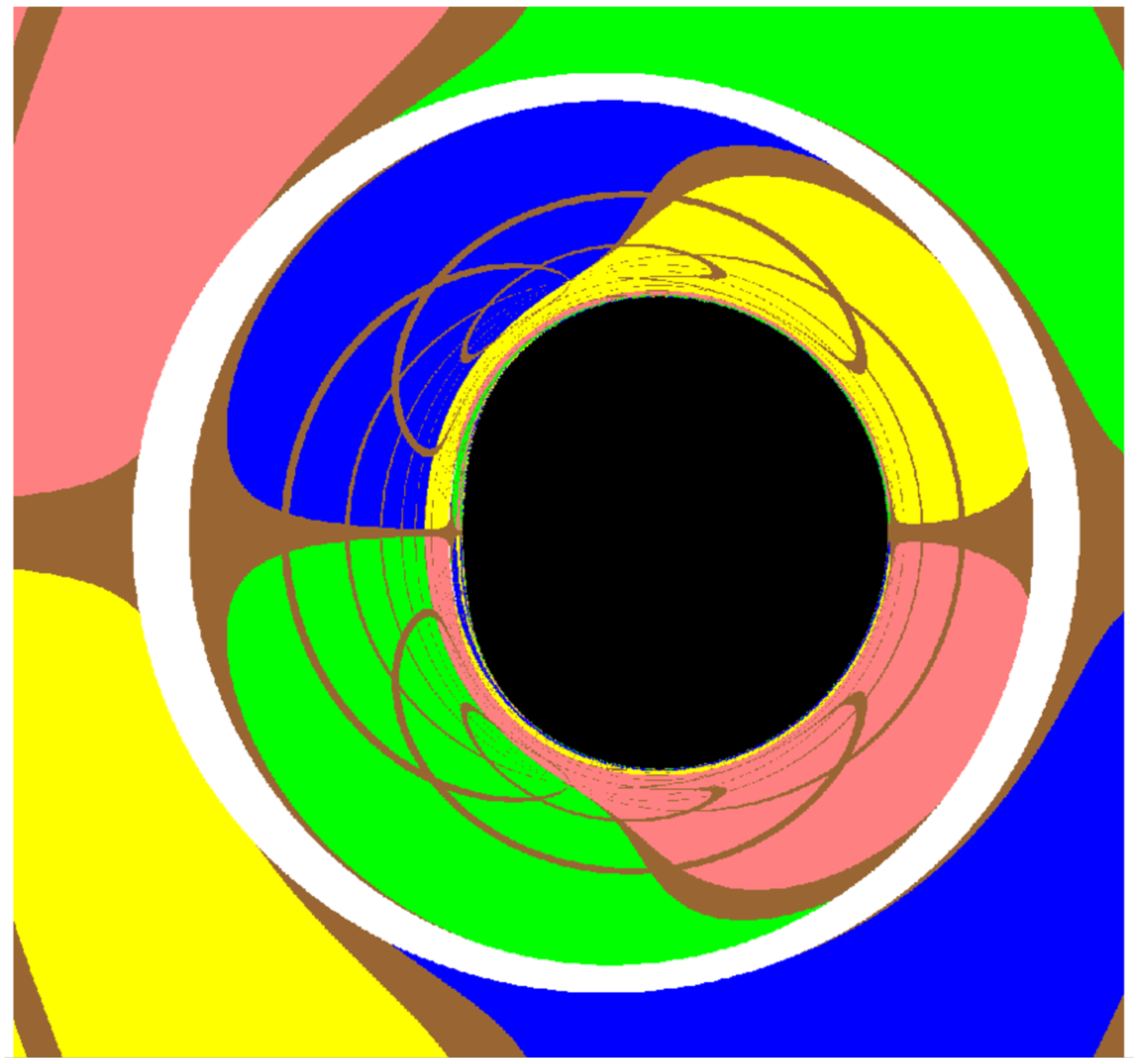}}
\caption{(a)The image of the background light source, in which there is no black hole at the center of the light sphere, seen by the observer with the backward ray-tracing method. (b)The image of Schwarzschild black hole seen by the observer. (c)The image of Kerr black hole with spin parameter $a=0.998M$ seen by the observer with the inclination angle $\theta_{0}=0$. (d)The image of Kerr black hole with spin parameter $a=0.998M$ seen by the observer with the inclination angle $\theta_{0}=\pi/2$\cite{sha18,sw}.}
\label{schk}}
\end{figure}

Fig.\ref{schk}(a) shows the image of the background light source, in which there is no black hole at the center of the light sphere, seen by the observer with the backward ray-tracing method. Fig.\ref{schk}(b) shows the image of Schwarzschild black hole seen by the observer. The black disk in the center is Schwarzschild black hole shadow; the bright colored image is the image of the background light source. And the white circle is the Einstein ring which is the image of the white reference spot on light sphere. Obviously, the image shows the curve of space caused by black hole and the effect of gravitational lensing. Fig.\ref{schk}(c) shows the image of Kerr black hole with spin parameter $a=0.998M$ seen by the observer with the inclination angle $\theta_{0}=0$. The dragging effect of the black hole can be clearly seen from the image around black hole shadow. Fig.\ref{schk}(d) shows the image of Kerr black hole with spin parameter $a=0.998M$ seen by the observer with the inclination angle $\theta_{0}=\pi/2$, it exhibits a ``D" shaped shadow.

\subsection{Chaotic shadows}

If the perturbation of a Schwarzschild or Kerr black hole is large enough, some photon motions in the non-integrable photon motion system could become chaotic. The chaotic photon motions are very sensitive to the initial conditions and present the intrinsic random in system. The black hole shadow will have a huge change caused by chaotic photon motions, it is very necessary to study the new essential features on the patterns and structures of black hole shadow.

\begin{figure}
\subfigure[]{ \includegraphics[width=3.9cm ]{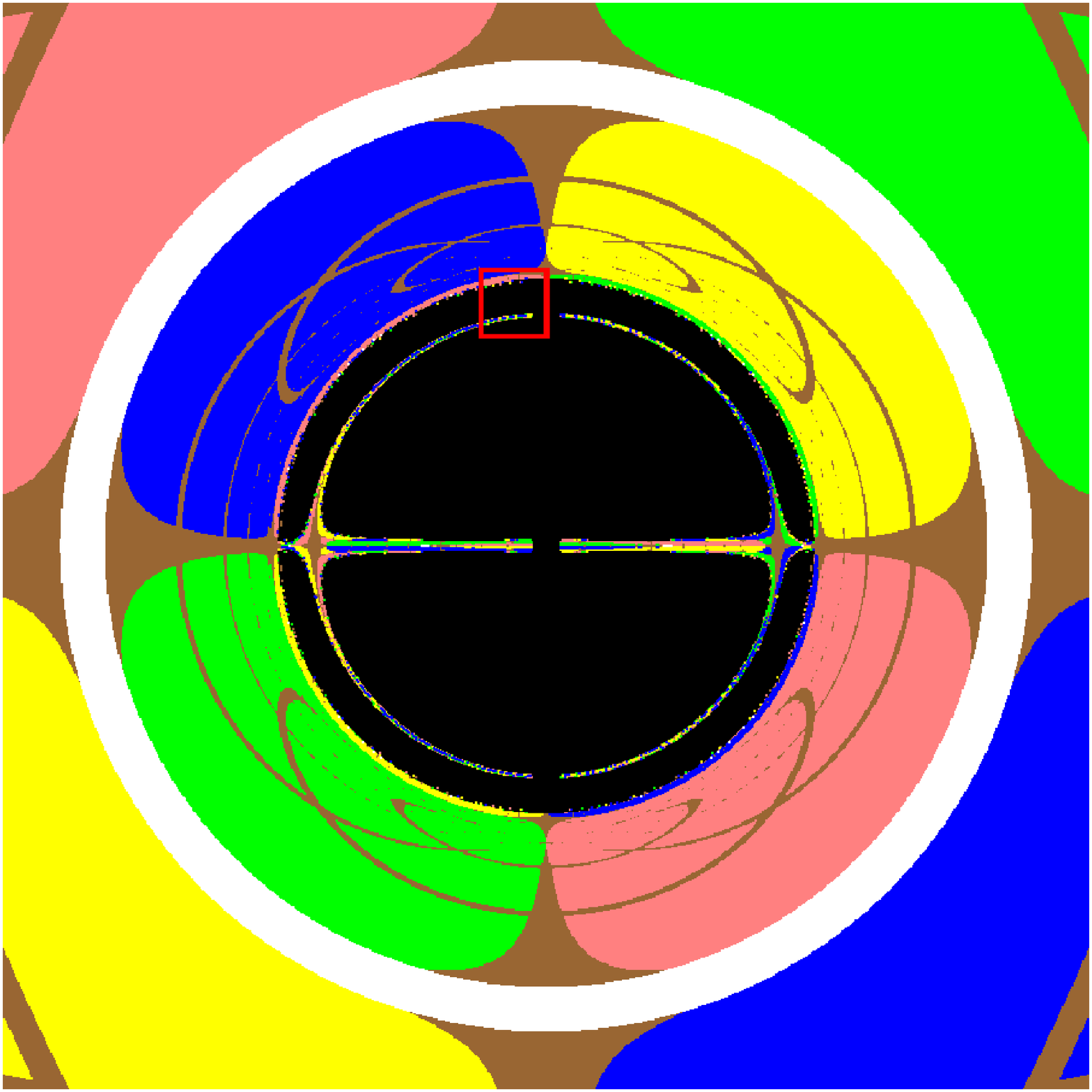}}\subfigure[]{ \includegraphics[width=3.9cm]{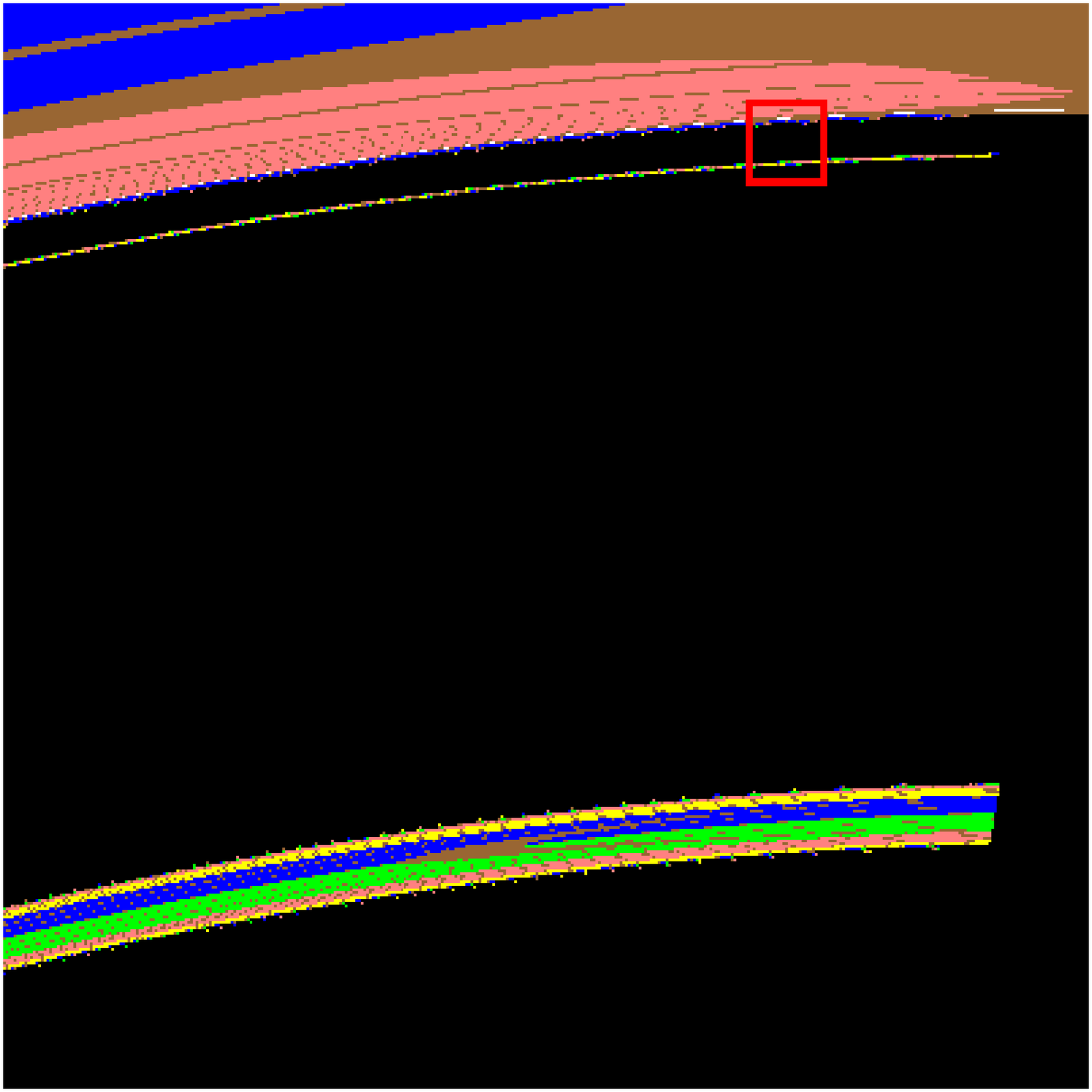}}\\
\subfigure[]{ \includegraphics[width=3.9cm ]{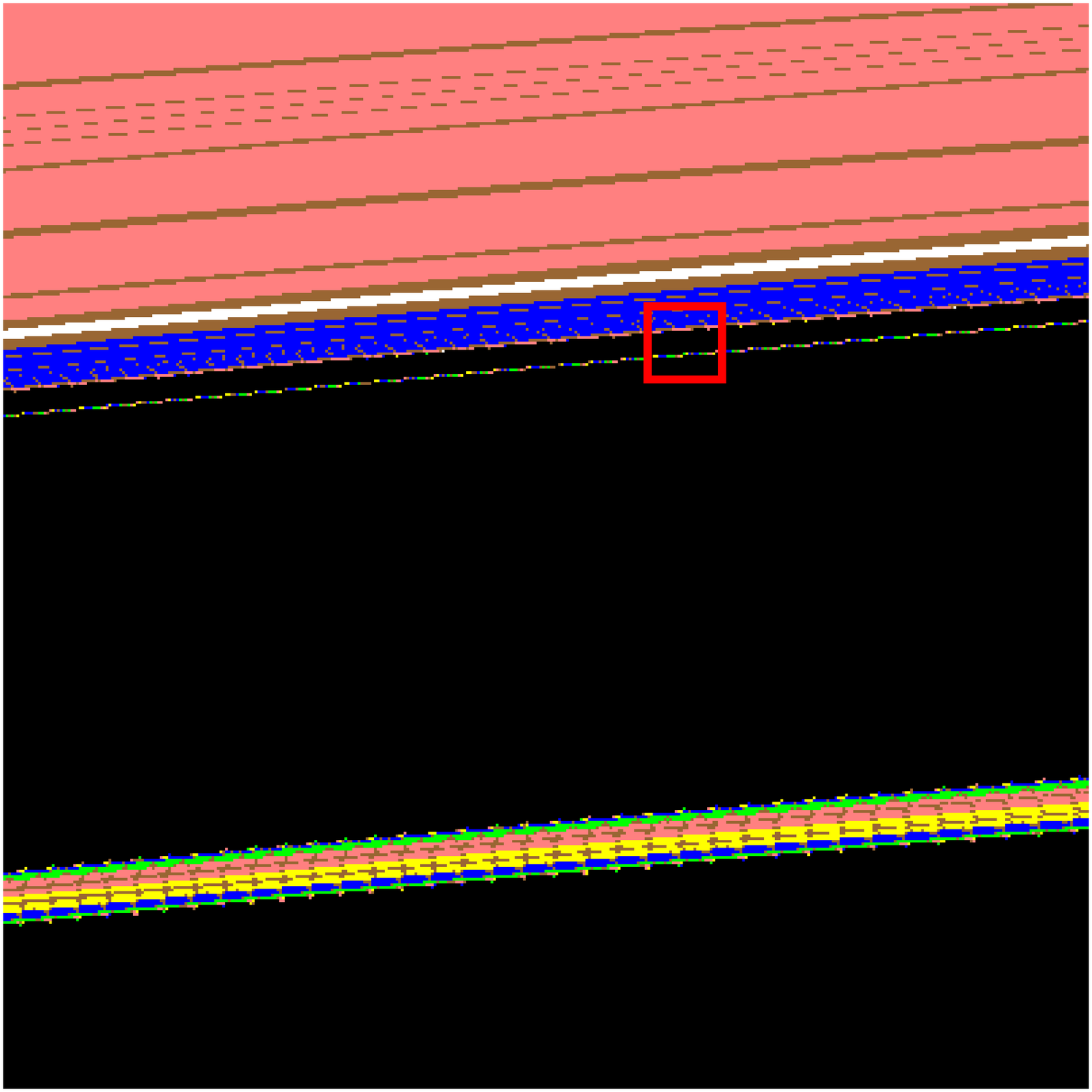}}\subfigure[]{ \includegraphics[width=3.9cm ]{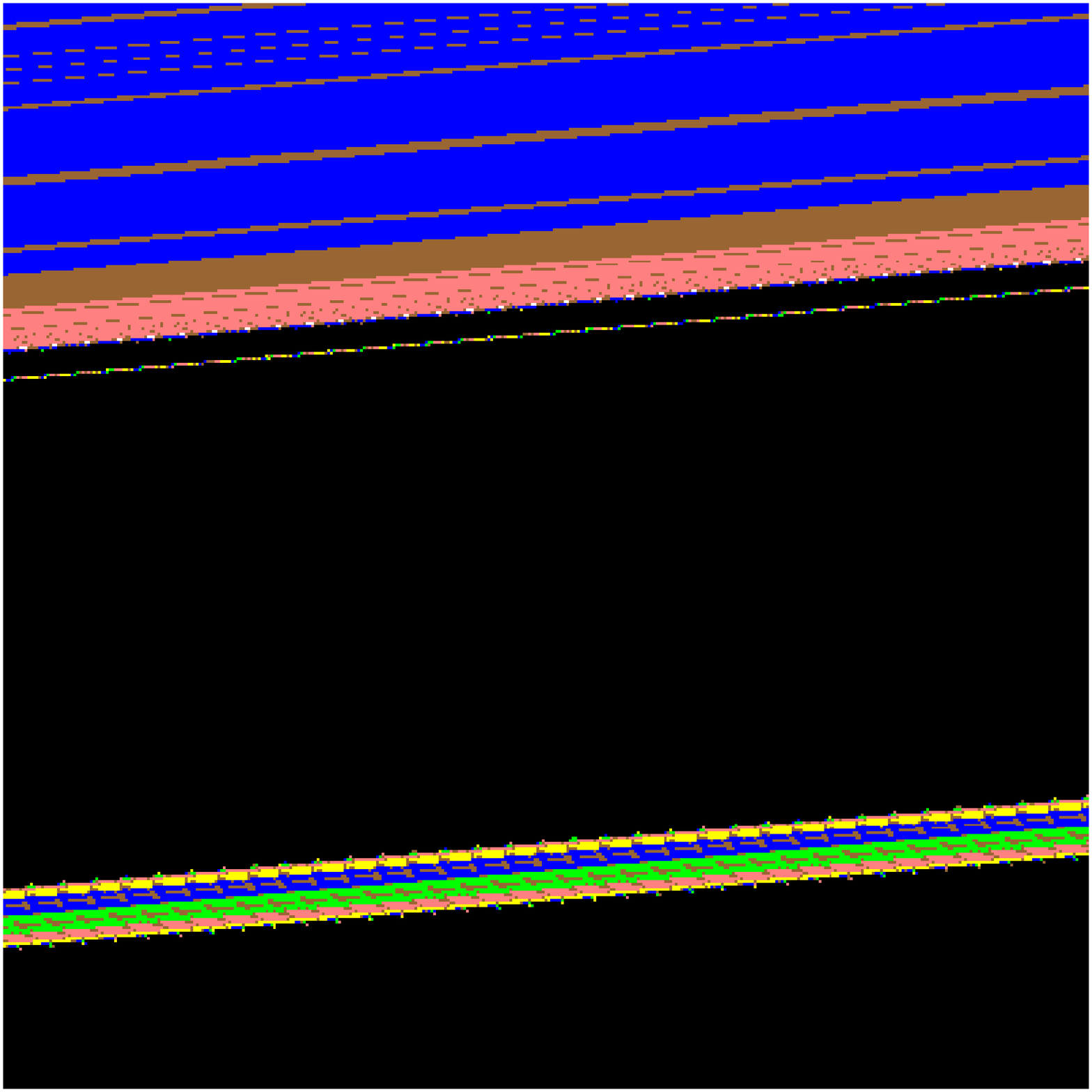}}
\caption{The fractal structures in the shadow of a compact object with a magnetic dipole\cite{my}.}
\label{fx}
\end{figure}

We have studied numerically the shadows of a compact object with magnetic dipole through the technique of backward ray-tracing\cite{my}. The presence of magnetic dipole makes the dynamical system of photon motion non-integrable, and affects sharply the shadow of the compact object. The Bonnor black dihole shadow is a concave disk with eyebrows with the larger magnetic dipole parameter, shown in Fig.\ref{fx}(a). We find the boundary of shadow in the red box has some layered structures. Amplifying further this shadow boundary, some similar layered structures are found as shown in Fig.\ref{fx}(b). We zoom in on the region within the red box in Fig.\ref{fx}(b) to got Fig.\ref{fx}(c), and continue to zoom in on the region within the red box in Fig.\ref{fx}(c) to got Fig.\ref{fx}(d). They are self-similar fractal structures caused by the chaotic photon motions. Actually, the main feature of chaotic shadow is the self-similar fractal structures. Our results show that the chaotic motions of photons could yield the novel patterns for black hole shadow.

\begin{figure}
\center{\includegraphics[width=7cm ]{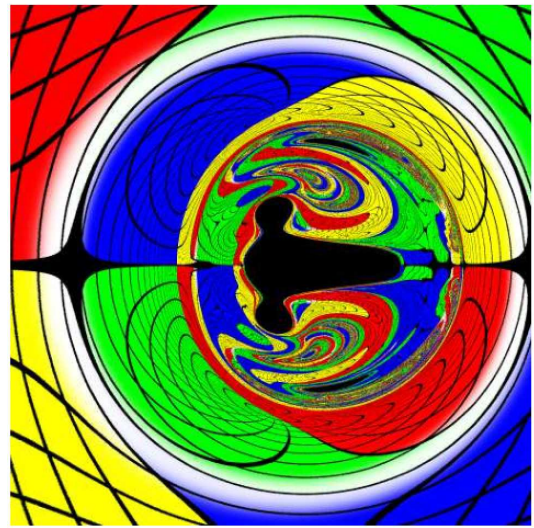}
\caption{The shadow of Kerr black hole with scalar hair\cite{sw}.}
\label{khair}}
\end{figure}

P. Cunha \cite{sw} studied the shadow of Kerr black holes with scalar hair. The metric functions are determined numerically by solving Einstein field equations and Klein-Gordon equations for scalar fields. The shadow of Kerr black hole with scalar hair is shown in Fig.\ref{khair}\cite{sw}. One can find the main shadow is shaped like a hammer, and there are many smaller eyebrow-like shadows on the top and bottom of the main shadow, which have self-similar fractal structures.

Furthermore, many scholars have studied various black hole shadows with chaotic characteristics, such as Kerr Black hole shadows in Melvin magnetic field\cite{scc}, Kerr black hole shadow casted by polarized lights\cite{pe}, the shadows of a Schwarzschild black hole surrounded by a Bach-Weyl ring\cite{mbw}, the shadows of non-Kerr rotating compact object with quadrupole mass moment\cite{sMN}, binary black hole shadows\cite{zhengwen23,sha18,binary,bsk}, and so on.

\subsection{Binary black hole shadows}

Nowadays, several gravitational waves events have been detected by LIGO-Virgo-KAGRA Collaborations, which are caused by binary black hole(BBH) merger\cite{js19,js20,js21,gw1,gw2,gw3,gw4} or by binary neutron star merger\cite{gw5}. It confirmed the existence of BBH in the Universe, and BBHs are expected to be common astrophysical systems.

\begin{figure}
\center{\includegraphics[width=8cm ]{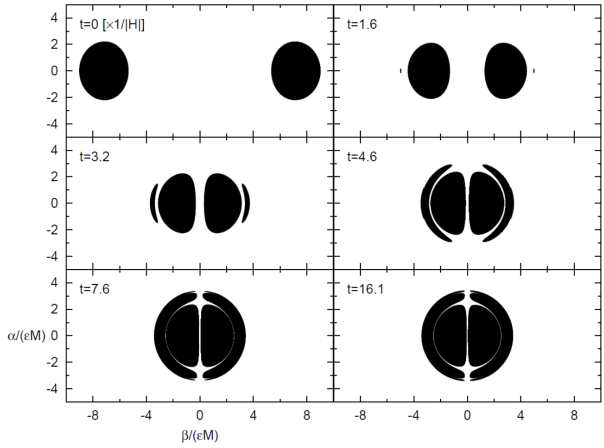}
\caption{The variation of black hole shadows with time $t$ during the collision of two equal mass black holes\cite{zhengwen23}.}
\label{collid}}
\end{figure}

While the shadow caused by an isolated black hole has been introduced above, what does a binary black hole looks like? D. Nittathe considered the Kastor-Traschen cosmological multiblack hole solution that is an exact solution describing the collision of maximally charged black holes with a
positive cosmological constant, and computed the shadows of the colliding of the two black holes\cite{zhengwen23}. Fig.\ref{collid} shows the variation of black hole shadows with time $t$ during the collision of the two equal mass black holes. At $t=0$, the two black holes are far enough apart that their shadows don't merge. However, one can find that each shadow is a little bit elongated due to the interaction between the two black holes. At $t=3.2$ (and even at $t=1.6$), the eyebrowlike shadows appear around the main shadows. At $t=7.6$, the eyebrow-like structures grow and the main shadows come close each other. The fractal structures emerge on the boundary of the main shadow and eyebrowlike shadows.
One can find the two shadows do not merge, and photons can go through between the main shadows. Even at $t=16.1$,  there still remains a region where photons can go through between the main shadows \cite{zhengwen23,zhengwen24}.

\begin{figure}
\subfigure[]{\includegraphics[width=4.3cm ]{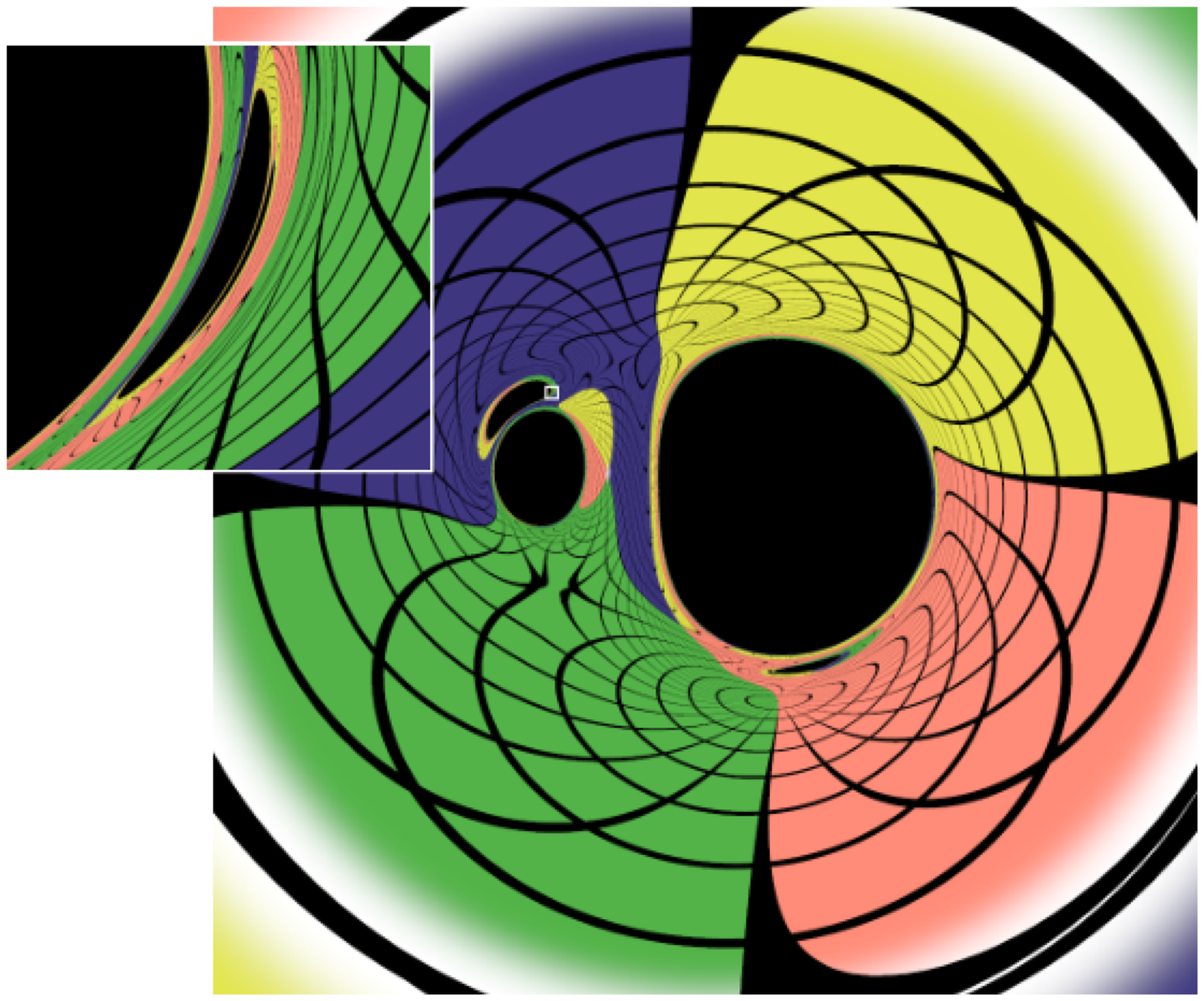}}\subfigure[]{\includegraphics[width=3.6cm ]{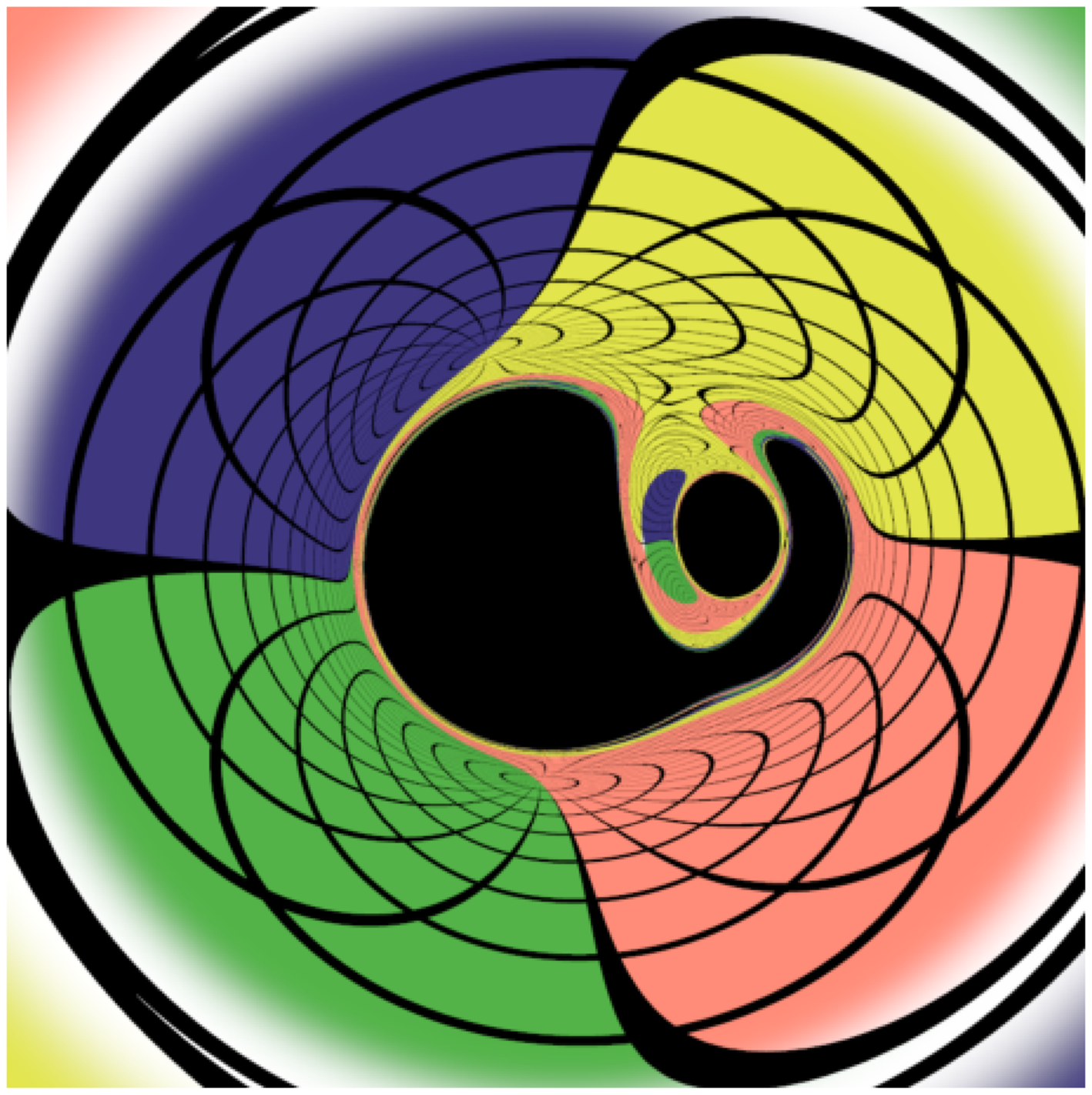}}
\caption{The shadows of BBH with the mass ratio $m_{1}/m_{2}=3$ and the spin parameters $a_{1}=0.7$ and $a_{2}=0.3$ respectively. The orbital angular momentum points out of the page in figure (a), and points up in figure (b)\cite{sha18}.}
\label{sshd}
\end{figure}

A. Bohn used the simulation of binary black holes inspiral space-time in Taylor's work\cite{taylor} to research the shadows of BBHs systems of two non-rotating black holes and two rotating black holes\cite{sha18}. Fig.\ref{sshd} shows the shadows of BBH with the mass ratio $m_{1}/m_{2}=3$ and the spin parameters $a_{1}=0.7$ and $a_{2}=0.3$ respectively\cite{sha18}. The orbital angular momentum points out of the page in Fig.\ref{sshd}(a), and points up in Fig.\ref{sshd}(b). From Fig.\ref{sshd}, one also can find the main shadows and eyebrowlike shadows with the fractal structures.

\begin{figure}
\includegraphics[width=7cm ]{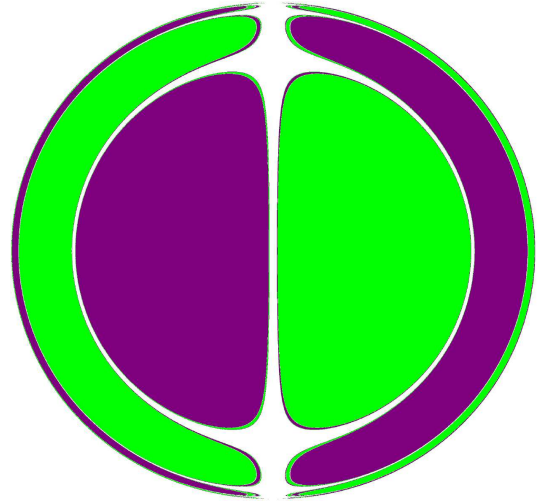}
\caption{The shadow of Majumdar-Papapetrou binary black hole\cite{binary}. The green regions represent the shadow of one black hole of Majumdar-Papapetrou BBH and the purple regions represent the shadow of the other black hole.}
\label{mp}
\end{figure}

J. O. Shipley\cite{binary} researched the shadows of Majumdar-Papapetrou binary black hole that is a solution of two extremally charged black holes in static equilibrium, in which gravitational attraction and electrostatic repulsion are in balance. They explained the eyebrowlike shadows form because lights bypass one black hole of BBH and enter the other. Fig.\ref{mp} shows the shadow of Majumdar-Papapetrou binary black hole\cite{binary}, in which the green regions represent the shadow of one black hole of Majumdar-Papapetrou BBH and the purple regions represent the shadow of the other black hole.

\begin{figure}
\subfigure[]{\includegraphics[width=4cm ]{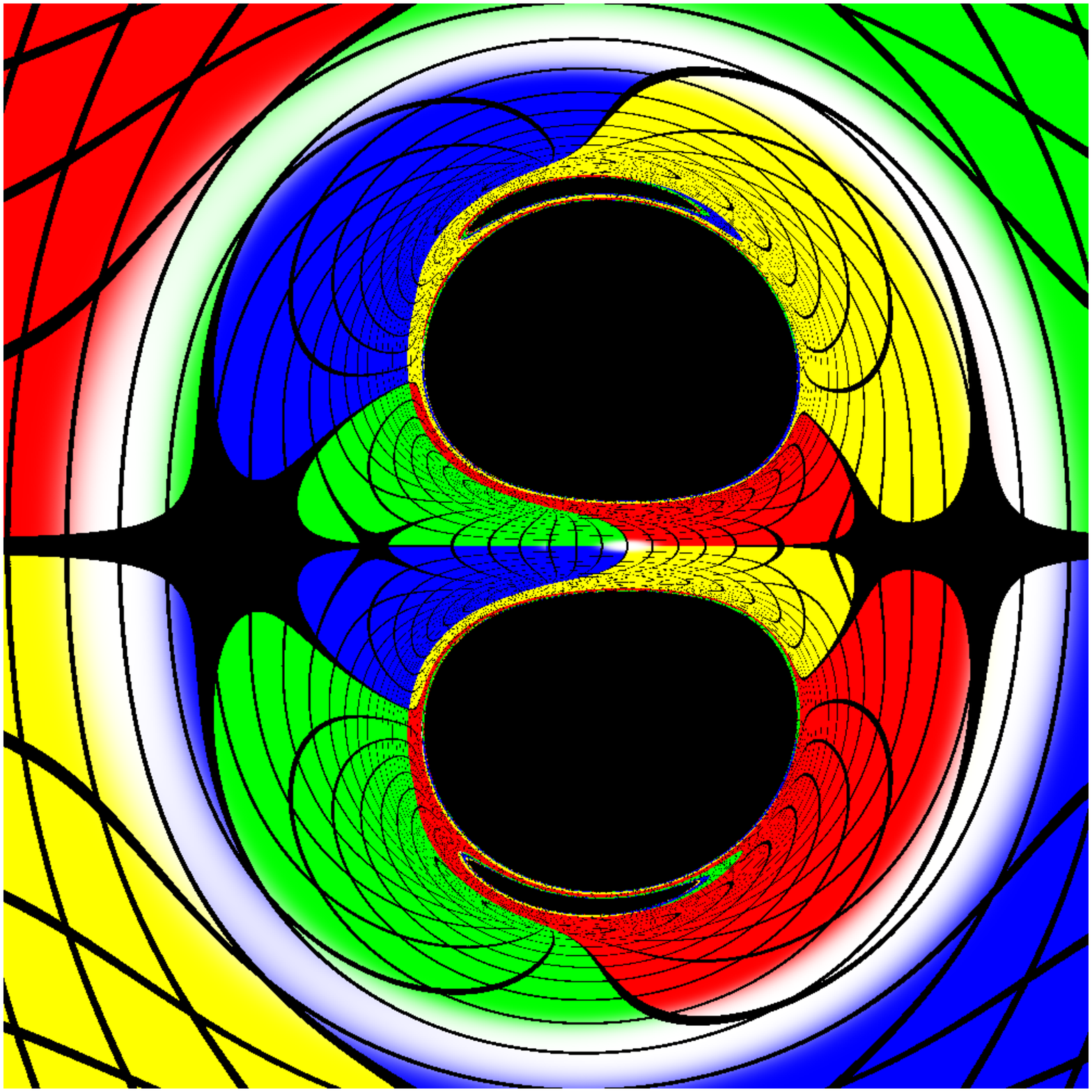}}\subfigure[]{\includegraphics[width=4cm ]{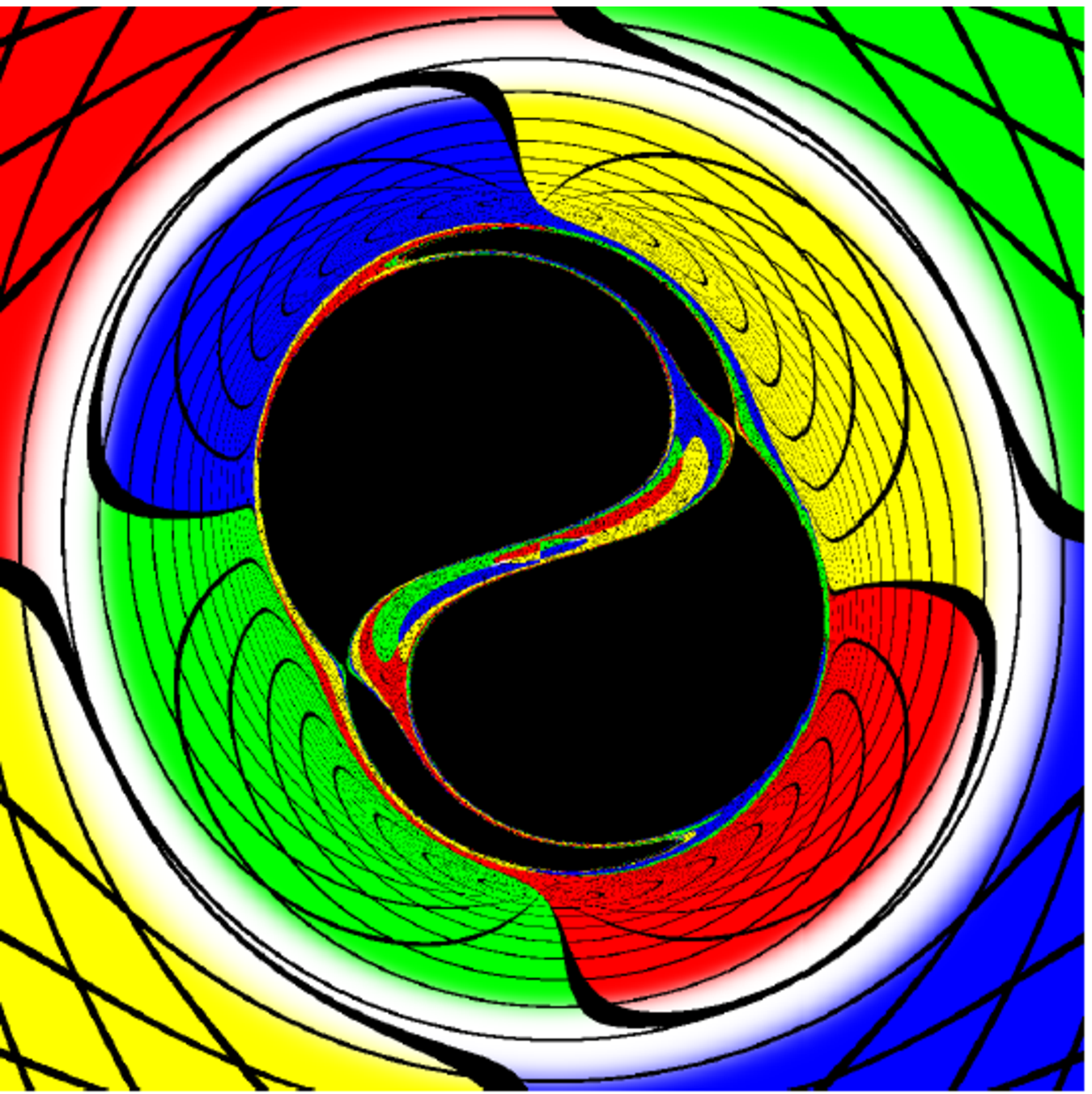}}
\caption{The shadows of double-Kerr black holes with equal spins (figure (a)) and with opposite spins (figure (b))\cite{bsk}.}
\label{dk}
\end{figure}

P.V.P. Cunha\cite{bsk} researched the shadows of double-Schwarzschild and double-Kerr black holes which are separated by a conical singularity. Fig.\ref{dk} shows the shadows of double-Kerr black holes with equal spins (figure (a)) and with opposite spins (figure (b))\cite{bsk}.

\subsection{The invariant phase space structures}
\begin{figure}
\subfigure[]{ \includegraphics[width=4cm ]{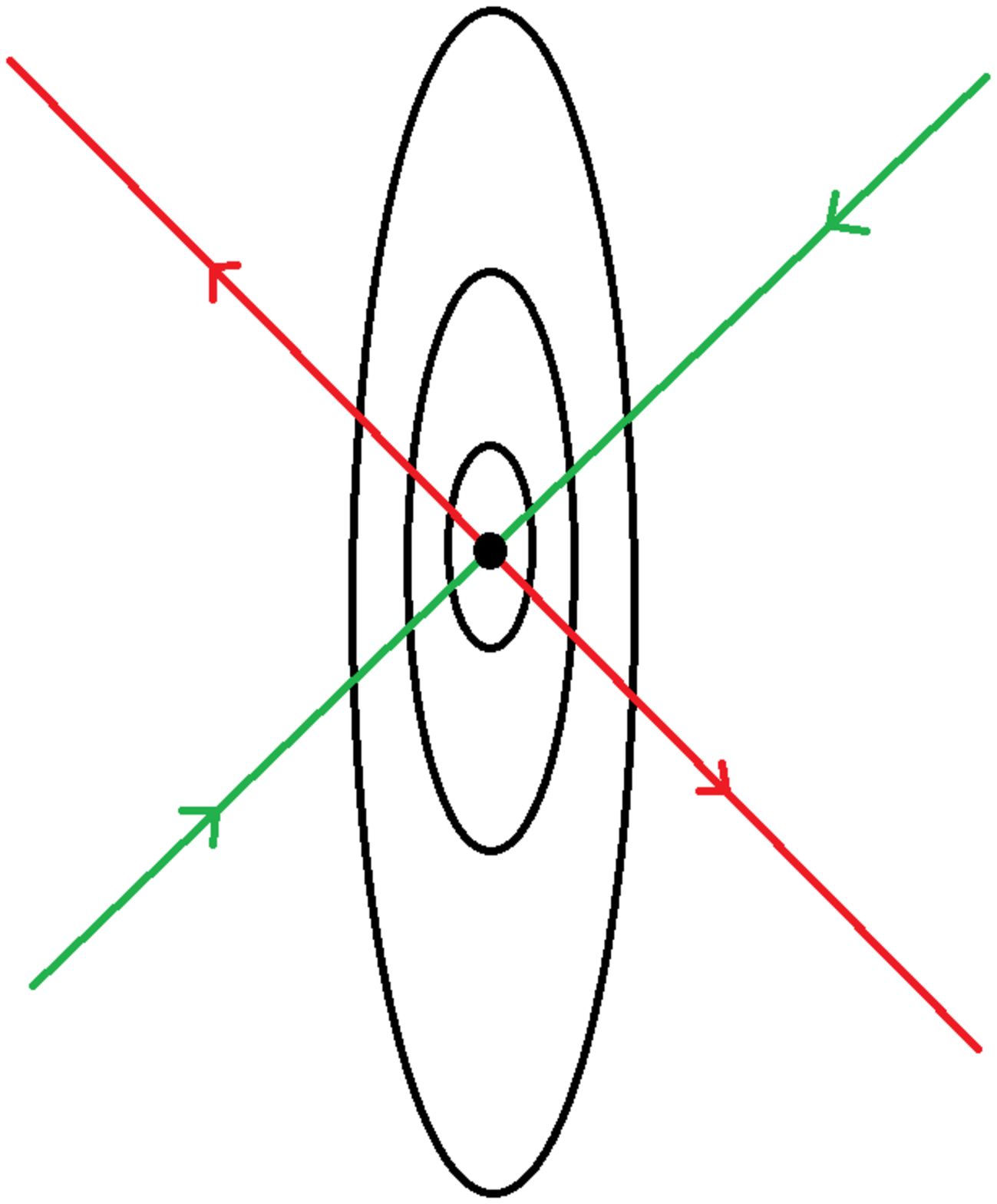}}\subfigure[]{ \includegraphics[width=4cm]{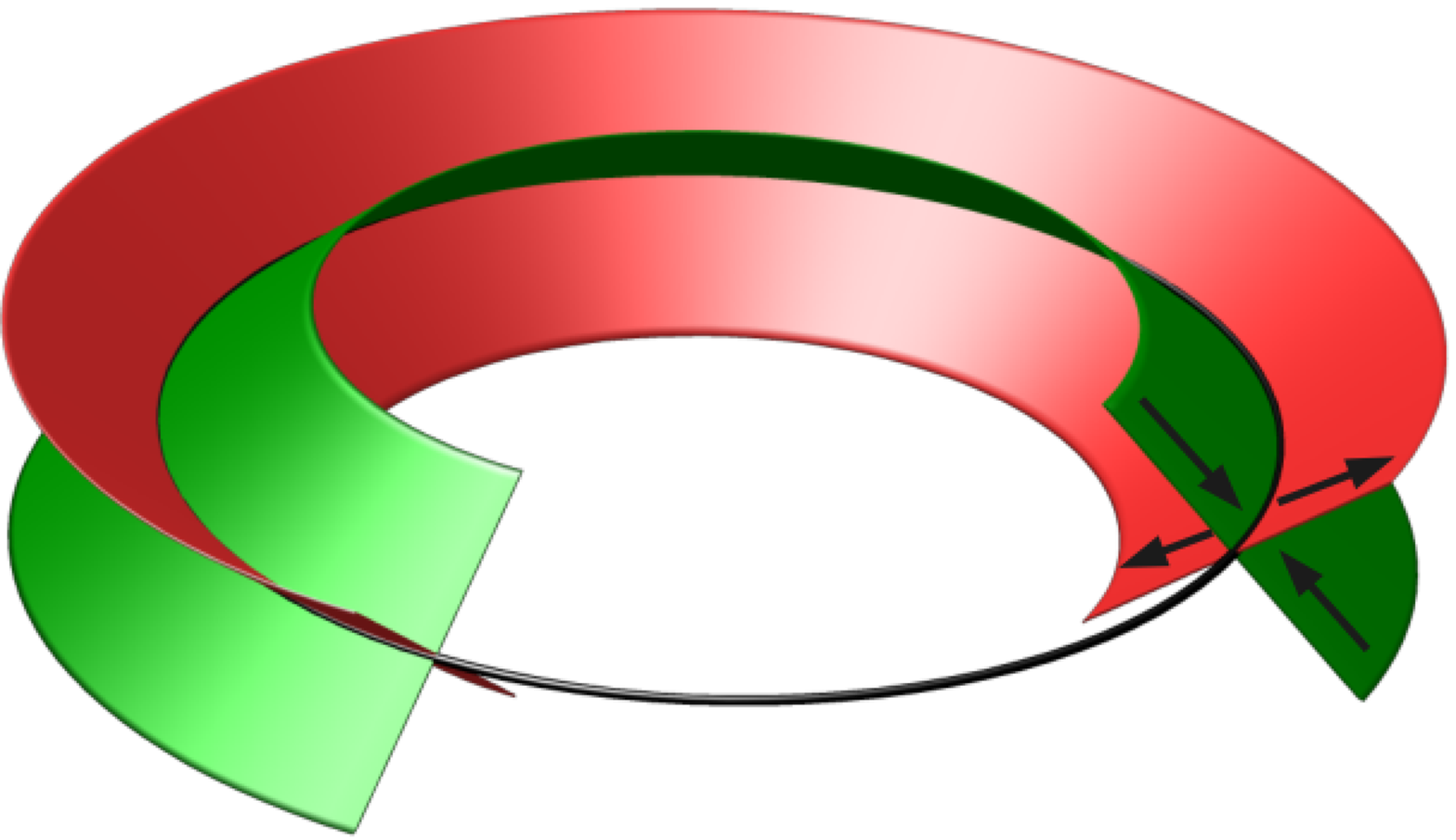}}
\caption{(a)The fixed point(black dot), the periodic orbits(the center manifold, black circles), the stable invariant manifold(green lines) and the unstable invariant manifold(red lines). (b)Stable (green) and unstable (red) manifolds of a periodic orbit (black circle)\cite{BI}.}
\label{lx}
\end{figure}

The invariant phase space structures are one of important features for dynamical systems, which are applied extensively in the design of space trajectory for various of spacecrafts, so they are also applicable to the study of light propagation. J. Grover's investigation\cite{BI} shows that the invariant phase space structures play an important role in the emergence of shadow of Kerr black holes with scalar hair. We also studied the invariant phase space structures in the space-time of the compact object with magnetic dipole\cite{my}, and found they are associated with the concave shadow with eyebrows (Fig.\ref{fx}(a)). The invariant phase space structures include fixed points, periodic orbits and invariant manifolds. The fixed points are the simplest invariant phase space structure, which satisfy the following conditions:
\begin{eqnarray}
\label{bdd}
\dot{x}^{\mu}=\frac{\partial H}{\partial p_{\mu}}=0,\;\;\;\;\;\;\;\;\;\;\;\;\;\;
\dot{p}_{\mu}=-\frac{\partial H}{\partial x^{\mu}}=0,
\end{eqnarray}
where $q^{\mu}=(t,r,\theta,\varphi)$ and $p_{\nu}=(p_{t},p_{r},p_{\theta},p_{\varphi})$. In black hole space-time, the photon circular orbits on the equatorial plane named light rings are fixed points ($r_{lr},\pi/2,0,0$) for the photon motion \cite{BI,fpos2}. By linearizing the equations (\ref{bdd}), one can obtain
\begin{eqnarray}
\label{xxh}
\mathbf{\dot{X}}=J\mathbf{X},
\end{eqnarray}
where $\mathbf{X}=(q^{\mu},p_{\nu})$ and $J$ is the Jacobian. The periodic orbits and invariant manifolds can be determined by the eigenvalues $\mu_{j}$ of the Jacobian $J$. The invariant manifolds are invariant under the dynamics, in which there is no trajectory can cross the invariant manifolds. If the real part of the eigenvalue of the Jacobian $J$ in the linearized equation (\ref{xxh}) is less than zero (${\rm Re}(\mu_{j})<0$), the corresponding eigenspace is unstable invariant manifold; if ${\rm Re}(\mu_{j})>0$, the corresponding eigenspace is stable invariant manifold; if ${\rm Re}(\mu_{j})=0$, the corresponding eigenspace is center manifold. Points in unstable (stable) invariant manifold exponentially approach the fixed points in backward (forward) time. According to Lyapunov central theorem, each purely imaginary eigenvalue gives rise to a one parameter family $\gamma_{\epsilon}$ of periodic orbits, the center manifold, which is also known as Lyapunov family \cite{BI} and the orbit $\gamma_{\epsilon}$ collapses into a fixed point as $\epsilon\rightarrow0$. Fig.\ref{lx}(a) exhibits the the fixed point(black dot), the periodic orbits(the center manifold, black circles), the stable invariant manifold(green lines) and the unstable invariant manifold(red lines). In addition, the periodic orbits also have their own stable and unstable manifolds. As shown in Fig.\ref{lx}(b), the green(red) surfaces represent the stable(unstable) manifolds of the periodic orbit (black circle), which exponentially approach the periodic orbit in backward (forward) time\cite{BI}.

\begin{figure}
\includegraphics[width=7cm ]{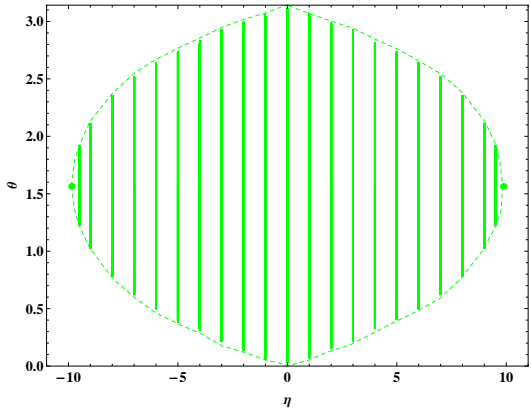}
\caption{Light rings (dots) and the corresponding family of periodic Lyapunov orbits (solid line) in the space-time of Bonnor black dihole\cite{my}.}
\label{zqt}
\end{figure}

The photon motion system in the space-time of the compact object with magnetic dipole, there are two fixed points, the counterclockwise and clockwise rotation light rings, shown in Fig.\ref{zqt}(green dots)\cite{my}. The solid green lines denote a family of periodic Lyapunov orbits arising from the two light rings, which can be parameterized by impact parameter $\eta$ shown in Fig.\ref{zqt}. Actually, these periodic Lyapunov orbits are the fundamental photon orbits which make up the photon sphere, so they are responsible for determining the boundary of black hole shadow.

\begin{figure}
\subfigure[]{\includegraphics[width=6.5cm ]{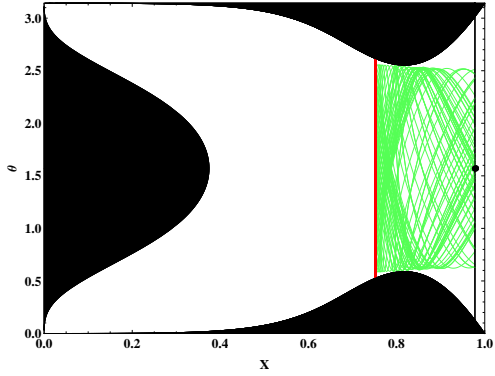}} \subfigure[]{\includegraphics[width=6.5cm ]{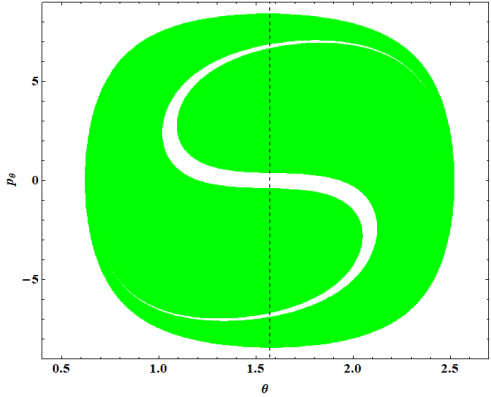}}
\subfigure[]{\includegraphics[width=6cm ]{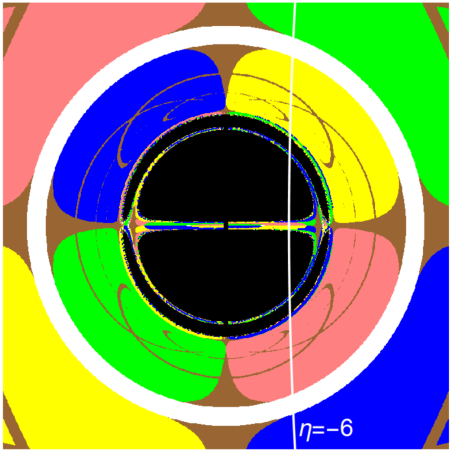}}
\caption{(a)Projection of the unstable invariant manifold (green lines) associated with the periodic orbit for $\eta$ =-6(red line)\cite{my}. The dark region are the forbid region of photon and the black dot represents the position of observer. (b)The Poincar\'{e} section(green) at the observers radial position in the plane ($\theta, p_{\theta}$) for the unstable invariant manifold of the periodic orbit with $\eta=-6$. (c)The shadow of Bonnor black dihole observed on the equatorial plane, which is marked with the line for $\eta=-6$.}
\label{sn}
\end{figure}

Fig.\ref{sn}(a) shows a projection of the unstable invariant manifold (green lines) associated with the periodic orbit for $\eta=-6$(red line) in the plane ($X,\theta$)\cite{my}, where $X$ is a compactified radial coordinate defined as $X=\sqrt{r^{2}-r_{h}^{2}}/(1+\sqrt{r^{2}-r_{h}^{2}})$\cite{BI}. From this figure, one can find the unstable invariant manifold builds a bridge between the photon sphere and the observer with $X\rightarrow1$(black dot). So the cross sections of unstable invariant manifolds at the observers radial position could reflect the shape of black hole shadow. Fig.\ref{sn}(b) shows the Poincar\'{e} section(green) at the observers radial position in the plane ($\theta, p_{\theta}$) for the unstable invariant manifold of the periodic orbit with $\eta=-6$. Moreover, there exist a white regions showing a ``S" shape which corresponds to photons move outside the unstable invariant manifold. The intersection of the dashed line $\theta=\frac{\pi}{2}$ with the unstable manifold denotes the trajectories with $\eta=-6$ which can be detected by the observer on the equatorial plane. The shadow of Bonnor black dihole observed on the equatorial plane is shown in Fig.\ref{sn}(c), which is marked with the line for $\eta=-6$. One can find the region of the unstable manifold that the dashed line $\theta=\frac{\pi}{2}$ passes through in Fig.\ref{sn}(b) corresponds to the part of black hole shadow that the line $\eta=-6$ passes through in Fig.\ref{sn}(c). And the anchor-like bright zone in black hole shadow in Fig.\ref{sn}(c) is originated from the top, middle and bottom parts of the ``S" shaped white region in the Poincar\'{e} section, Fig.\ref{sn}(b). It indicates that the formation of black hole shadow is related to the invariant manifolds of certain Lyapunov orbits near the fixed points.

\section{Summary}

In this review, we briefly described the formation and the calculation methods of black hole shadow. We first introduced the conception of black hole shadow, and the great importance of black hole shadow for the study of astrophysics and black hole physics. We also introduced the current works on various black hole shadows. Then we took Schwarzschild black hole shadow as an example to calculate the photon sphere radius $r_{\rm ps}$ and shadow radius $r_{sh}$, and explained that the boundary of shadow is determined by the photon sphere composed by the unstable photon circular orbits. Next, we took Kerr black hole shadow as an example to review the analytical calculation for black hole shadow, it has analytic expressions for shadow boundary due to the integrable photon motion system. And we introduced the fundamental photon orbits which can explain the patterns of black hole shadow shape. Finally, we review the numerical calculation of black hole shadows with the backward ray-tracing method, and introduce some chaotic black hole shadows with self-similar fractal structures casted by chaotic lensing. Since LIGO-Virgo-KAGRA Collaborations have detected the gravitational waves from the merger of binary black hole(BBH), we introduced a couple of shadows of BBHs. They all have the eyebrowlike shadows around the main shadows with the fractal structures. We computed the invariant phase space structures of photon motion system in black hole space-time, and explained the formation of black hole shadow is related to the invariant manifolds of certain Lyapunov orbits near the fixed points.

We hope the black hole shadows have obtained by theoretical calculation could provide some theoretical templates for the future astronomical observations announced by upgraded Event Horizon Telescope and BlackHoleCam. And then the study of black hole shadow could promote the development of black holes physics and verify various gravity theories.

\section{\bf Acknowledgments}

This work was supported by the National Natural Science Foundation of China under Grant No. 12105151, 11875026, 11875025 and 12035005£¬ and the Shandong Provincial Natural Science Foundation of China under Grant No. ZR2020QA080.

\end{document}